\documentclass[12pt,superscriptaddress]{article} 
\usepackage{latexsym}
\usepackage{graphics}
\usepackage{epsfig}
\usepackage[mathscr]{eucal}
\usepackage{amscd}
\usepackage{amssymb}
\usepackage{amsmath}

\newcommand{\re}{\mathbb R}
\newcommand{\bZ}{Z}
\newcommand{\Hom}{H} 

\newcommand{\bN}{\mathbb N} 
\newcommand{\nN}{\mathfrak N}

\newcommand{\nin}{\notin}
 
\newcommand{\vl}{\mathsf v}

\newcommand{\cM}{\mathsf M}

\newcommand{\cO}{\mathcal {O}}

\newcommand{\cI}{\mathcal I}

\newcommand{\fut}{\mathrm{Fut}}
\newcommand{\past}{\mathrm{Past}}
\newcommand{\ifut}{\mathrm{IFut}}
\newcommand{\ipast}{\mathrm{IPast}}

\newcommand{\alex}{\mathsf I}

\newcommand{\vol}{\mathrm{vol}}

\newcommand{\can}{A} 
  
\newcommand{\bA}{\mathcal A} 
\newcommand{\cP}{\mathcal P}
\newcommand{\hv}{\mathsf n} 
\newcommand{\thv}{\tilde \hv}
 
\newcommand{\bB}{ B} 

\newcommand{\al}{\alpha} 
\newcommand{\cT}{{{\mathsf T}}}
 
\newcommand{\be}{\begin{equation}}
\newcommand{\ee}{\end{equation}}

\newcommand{\pd}{\partial} 
\newcommand{\rc}{ C} 
 
\newcommand{\bM}{\mathbb M}
 
\newcommand{\cN}{\mathcal N}
\newcommand{\bI}{\mathcal I}
\newcommand{\mbs}{\mathbb S} 
\def\maximal{inextendible}

\def\ia{inextendible{ } antichain} 
\def\ias{inextendible{ } antichains} 
\def\ta{thickened{ } antichain} 
 
\def\hpt{Hauptvermutung}

\newcommand{\limit}{\mathrm {Lt}} 

\newcommand{\tvl}{\widetilde \vl}

\newcommand{\caV}{\mathcal V}
\newcommand{\cL}{\mathcal L}
\newcommand{\ld}{\lambda} 
\newtheorem{theorem}{Theorem}
\newtheorem{lemma}{Lemma}

\newtheorem{corollary}{Corollary}
\newtheorem{claim}{Claim}
\newcommand{\bproof}{\setlength{\parindent}{0mm}{\bf Proof{~~}}}
\newcommand{\eproof}{\hfill $\Box$\setlength{\parindent}{5mm}}
\newcommand{\diam}{\mathrm{diam}}  
 
\newcommand{\ef}{f}
\newcommand{\tef}{\widetilde{f}} 
\newcommand{\pdsz}{\pd \overline {\sigma_s(\vl)}}
\newcommand{\pdszt}{\pd \overline {\sigma_s(\vl_2)}}

\newcommand{\mm}{\mathrm m}
\newcommand{\NI}{\cN(\bI)}
\newcommand{\NP}{\cN(\cP)}
\newcommand{\hb}{\hat b}
\newcommand{\hbq}{\hat b_q} 

\newcommand{\cB}{\mathfrak B}
\newcommand{\nb}{\mathsf N}
\newcommand{\inte}{\mathrm {int}}
\newcommand{\mS}{\mathcal S} 
\newcommand{\cx}{\mathfrak r}
\newcommand{\cnn}{\mathsf N}
\newcommand{\cnnn}{\mathsf N'} 
\newcommand{\sy}{\mathsf y} 
\newcommand{\fE}{\mathfrak F}
\newcommand{\jj}{\mathsf J} 
\newcommand{\pc}{\kappa }
\newcommand{\mF}{\mathsf F}
\newcommand{\pcf}{\kappa^{\mathsf F}} 
\newcommand{\ind}{i}
\newcommand{\nsf}{\mathsf n}

\newcommand{\dimm}{\mathfrak n}
\newcommand{\Dimm}{\mathsf {|M|}} 
\newcommand{\pprec}{\prec\prec}
\newcommand{\cF}{\mathcal F}
\newcommand{\pP}{\mathsf P} 
\newcommand{\ssigma}{\Lambda}
\newcommand{\tf}{{\widetilde f}}
\newcommand{\tS}{{\widetilde S}} 
\newcommand{\vcrit}{\vl_{\mathrm{crit}}}
\title{ On Recovering Continuum Topology  \\ from a Causal Set}   

\author{Seth Major${}^1$, David Rideout${}^2$ and Sumati Surya${}^3$
  \\  
${}^1$Department of Physics, Hamilton College  Clinton, NY 13323,
  U.S.A. \\ 
${}^2$Blackett Laboratory, Imperial College, London, SW7 2AZ, UK \\
${}^3$Raman Research Institute, Bangalore, India \\ \\
} 
\begin{document}
\maketitle 

\begin{abstract} 
An important question that discrete approaches to quantum gravity must
address is how continuum features of spacetime can be recovered from
the discrete substructure.  Here, we examine this question within the
causal set approach to quantum gravity, where the substructure
replacing the spacetime continuum is a locally finite partial order. A
new topology on causal sets using ``thickened antichains'' is
constructed. This topology is then used to recover the homology of a
globally hyperbolic spacetime from a causal set which faithfully
embeds into it at sufficiently high sprinkling density. This implies a
discrete-continuum correspondence which lends support to the
fundamental conjecture or ``Hauptvermutung'' of causal set theory.
\end{abstract}

\section{Introduction}  

Spacetime discretisation is a common calculational device used to
regularise background dependent physics. Typically, the discretisation
is topologically trivial, with the spacetime continuum replaced by a
lattice which is regular in a preferred reference frame.  Physical
results are then obtained by taking this cut-off to zero.  Two
important issues which arise as a result of a naive spacetime
discretisation are already apparent in quantum field theory on
Minkowski spacetime. The first is the breaking of Poincare
invariance, and the second, the loss of global topological
information. Since discreteness is used only as a calculational tool,
these issues only pose practical limits on the discretisation, since
relevant physics is, by and large, recovered in the continuum limit.

However, both issues assume a more fundamental role in discrete
approaches to quantum gravity, in which the continuum is taken to
arise as an approximation, rather than as a limit.  Instead of
being a means to regulate the theory, spacetime discreteness is taken
to be fundamental, much like the atomicity of an apparently continuous
fluid.  The choice of the discrete building blocks  in a given
approach to quantum gravity then determines the manner in which 
these two issues manifest themselves. 
 
In many discrete approaches to quantum gravity, local Lorentz
invariance is explicitly broken, and much recent work has been devoted
to quantifying such violations.  For instance, in the case of modified
dispersion relations, threshold analyses demonstrate that current
astrophysical observations place severe constraints on cubic
modifications \cite{MDR}.  Significantly, in the causal set approach
to quantum gravity, no such violation occurs. This unique property
arises from the fact that the continuum approximation of the theory
obtains from a random process \cite{dhs}.

In this work, we address the question of how to recover 
continuum
topology from the discrete substructure within the framework of causal
set quantum gravity \cite{bometal}. In this approach, the spacetime
continuum is replaced by a causal set or causet which is a locally
finite partially ordered set. The continuum approximation of the
theory obtains from a ``{faithful embedding}'' $\Phi:\rc \rightarrow
M$ of the causet $\rc$ to a spacetime $(M,g)$, at spacetime
density $\rho$, i.e., $\Phi(\rc) \subset M$ is a high probability
random (Poisson) distribution of points on $M$ at density $\rho$ such
that the order relation in $\rc$ is mapped to a causal relation in
$(M,g)$.  A causet embedded in $(M,g)$ thus resembles a random
spacetime lattice, where the directed links between two points
indicate a causal relation.

A key conjecture of causal set theory, the ``Hauptvermutung'', states
that the continuum approximation of a causet is unique upto an
approximate isometry; thus, if $\Phi:\rc \rightarrow (M,g)$ is a
faithful embedding at density $\rho$, then $(M,g)$ is unique upto
isomorphisms at ``scales above $\rho^{-1}$''. While this has been
proved rigourously in the infinitely fine grained limit, $G \hbar
\rightarrow 0$ \cite{bommeyer}, there is evidence from calculations of
dimensions and geodesic distance that provide support for the
conjecture in the finite case \cite{meyer,bg}. For example, for a
causet $C$ which faithfully embeds into $d$-dimensional Minkowski
spacetime $\bM^d$, the ``Myrheim-Meyer'' dimension of an interval in
$C$ gives a good estimate of the continuum dimension
\cite{meyer}. Thus, if a causet $C$ embeds faithfully
into $\bM^{d_1}$ and $\bM^{d_2}$ at the same sprinkling density, then
$d_1 \sim d_2$.

It is therefore of interest to seek a correspondence between the
continuum topology and an appropriately defined topology on the causal
set. Such a correspondence would then imply that if a causet
faithfully embeds into two spacetimes with topology $M_1$ and $M_2$,
then there is an approximate homeomorphism $M_1 \sim M_2$. This would
also imply that $M_1$ and $M_2$ are homologous at scales larger than
$\rho^{-1}$.  In this work we make considerable progress in
establishing a correspondence between the homology of thickened
antichains which are special subcausets of $\rc$ and that of a
globally hyperbolic spacetime into which it faithfully embeds. Under
certain assumptions, this implies a homological version of the
Hauptvermutung.

The random nature of the sprinkled causal set makes the task of
finding a correspondence fairly non-trivial. In approaches using
simplicial methods like Dynamical Triangulations or Spin Foams, the
discrete structure can be taken to be a triangulation of the given
manifold, which though non-diffeomorphic to the continuum, by
construction, carries all continuum homological information
\cite{DT}. Conversely, an abstract simplicial complex is associated
with a given manifold only if it can be be mapped bijectively to a
triangulation of that space.  In causal set theory, however, this
connection is apparently more tenuous, since the discretisation does
not preserve continuum topological information in an obvious way.

However, a non-trivial partial order does possess sufficient structure
compared to the unordered set of points on a lattice, and hence admits 
non-trivial topologies \cite{stanley,posetbook}.  For example, a {\it
chain complex} on $C$ is constructed by mapping every $k$-element {\sl
chain}, or totally ordered subset, to a $k$-simplex, while the {\sl
interval topology} is constructed from sets which are 
analogs of the Alexandrov intervals in a spacetime
\cite{posetbook}. Indeed, it has been recently shown that a globally
hyperbolic spacetime is a so-called bicontinuous poset\footnote{In
particular, this means that the poset is not locally finite, and hence
not a causal set according to our definition.} whose interval topology
is the same as the manifold topology \cite{prakash}. Thus, partially
ordered sets admit very rich topological structures.

For a locally finite causal set, however, it is unclear how these
topologies relate to the continuum topology. An important first step
is to realise that the topology of the continuum approximation is too
rich and contains ``irrelevant'' information on scales below the
discreteness scale $\rho^{-1}$. Thus, the assumption of a fundamental
discreteness is incompatible with the requirement that there exist a
strict homeomorphism between the causal set topology and the continuum
topology. Physical significance cannot be attributed to continuum
structures of characteristic size smaller than $\rho^{-1}$ and hence
only macroscopic topological information, like homology or homotopy at
scales $\gg \rho^{-1}$ can be considered relevant to the causet\footnote{This
observation is true of any finitary topology \cite{finitary}}.

In this work we provide a prescription for constructing a map between
discrete and continuum homologies  for the class of causal sets that
faithfully embed into globally hyperbolic spacetimes.  The simplicial
complex we construct for the causal set uses the discrete analog of a
Cauchy hypersurface, i.e., an ``{\sl \maximal{} antichain}'' or a
maximal set of unordered points in the causal set \cite{mots}.
While an \maximal{} antichain, being endowed with only the trivial
topology, does not itself suffice to capture any continuum topological
information, it does inherit non-trivial structure from its embedding
in the larger causal set to which it belongs. As in \cite{mots},
starting with an \maximal{} antichain $\can$ we
define the   discrete analog of a spacetime volume ``thickened'' Cauchy
hypersurface or  thickened antichain 
\begin{equation} 
\label{thickening} 
\cT_{\hv}(\can) \equiv \{ \, p \, | \, \, | ( \{ p \}  \sqcup  \past(p)) 
\cap (\{ \can \} \sqcup \fut(\can))| \leq {\hv}\}, 
\end{equation} 
where ${\hv} \in \bN$, $|X|$ is the cardinality of a set $X$ and
$\past(p)$ and $\fut(p)$ are the past and future of $p$,
respectively. 
For finite ${\hv}$, every element in $\cT_{\hv}(\can)$ has a finite past in
$\cT_{\hv}(\can)$, and hence one can find the set of maximal or future-most
points in $\cT_{\hv}(\can)$.  A nerve simplicial complex is then
constructed from the shadows of the past sets of these maximal
elements onto $\can$, details of which will be described in the
following sections \cite{finitary}. Non-trivial overlaps of these
shadows implies a non-trivial simplicial complex from which one can
extract homological information.

On the face of it, there is nothing to suggest that the above
construction is more natural than the chain or interval
topologies. However, preliminary numerical simulations in 1+1
dimensions suggest the existence of a robust correspondence between
this causal set topology and the continuum topology
\cite{numerical}. In this work, using purely analytical tools, we show
that the continuum analogue of this construction in a globally
hyperbolic spacetime does indeed yield a nerve simplicial complex
which is not only homologous but homotopic to the spacetime.
Moreover, we show that there exists an isomorphism between the
homology groups of the discrete and continuum nerves, for causets
which faithfully embed into globally hyperbolic spacetimes.

We begin with some basic definitions in Section 2. We then
construct the nerve simplicial complex for the discrete and the
continuum cases in Section 3. In Section 4, we prove an important
continuum result. Namely, we show that the continuum nerve is a
simplicial complex which is homotopic to the spacetime manifold $M$
for a class of ``volume thickenings'' of a Cauchy surface. 
We make crucial use of a theorem due to de Rham and Weil on the nerve
of a locally finite convex cover.  Finally, we prove our main result
in Section 5, i.e., we show for a class of \ias{} in a faithful
embedding that with high probability the order-theoretic nerve
simplicial complex is homologically equivalent to the manifold for
sufficiently high sprinkling densities $\rho$. We summarise our
results in Section 6 and conclude with a discussion of some of the
open questions. Since this work is heavy with notation, the appendix
provides a table of the symbols used and their definitions.

\section{Preliminaries}

Here we set down some definitions and notations that we will need.

\noindent {\bf The Causal Set:} 

A {\sl causal set} $C$ is a set with an order relation
 $\prec$ which is  (i) Transitive, i.e., $x \prec y$ and $ y \prec z
 \Rightarrow x 
\prec z$),  (ii)  Irreflexive, i.e., $ x \nprec x $ 
and (iii) Locally finite, i.e. $|\past(x) \cap \fut(y)| <
\infty$ for any $x,y,z \in C$, where $\past(y)\!=\! \{x| x\prec y
\}$, $\fut(y)\!=\!\{x| y \prec x \}$ and $|A|$ denotes
set-cardinality.

These discrete analogs of the causal future/past sets of the
continuum do not include $x$
because of the irreflexive condition. Since such an inclusion will
find use in our constructions, we define the inclusive future and past
sets as $\ifut(x) \equiv x \cup \fut(x) $ and $\ipast \equiv x \cup
\past(x) $, respectively.

\noindent {\bf Causal Relations in the Continuum:} 

We will use the notation and results from \cite{HE}. $\alex^\pm(p)$
denotes the chronological future/past of an event $p$ and $\jj^\pm(p)$
its causal future/past. We will refer to the region spacelike to $p$,
by $S(p) \equiv M \backslash (\jj^+(p) \cup \jj^-(p))$.  The {\sl
Alexandrov interval} is defined to be the open set $\alex(p,q) \equiv
\alex^+(p) \cap \alex^-(q)$. The generalisation of these definitions
for sets is straightforward, as is the notation.  In a globally
hyperbolic spacetime $(M,g)$, $\overline{\alex(p,q)}$ is compact for
any $p,q \in M$. For $p, q \in M$ we will use the notation $p \prec
\prec q $, $p \prec q $ and $p \rightarrow q$ to denote a
chronological, causal and null relation, respectively. We will find
use for the result, $p \pprec q$, $q \prec r $ $\Rightarrow p \pprec
r$.

Let $F$ be a function which assigns to each event $x$ in $M$
an open set $F(x) \subset M$.  Then $F$ is said to be {\sl inner
continuous} if for any $x$ and any compact $K \subset F(x)$, there
exists a neighbourhood $U$ of $x$ with $K\subset F(y)$ for all $y \in
U$.  $F$ is said to be {\sl outer continuous} if for any $x$ and any
compact set $K \subset M \setminus {\overline {F(x)}}$, there exists a
neighbourhood $U$ of $x$ with $K \subset M \setminus {\overline {F(y)}}$
$\forall y \in U$.

In a time-orientable distinguishing spacetime, $(M,g)$ the sets
$\alex^+(x)$ and $\alex^-(x)$ are inner continuous \cite{HS}, but need
not be outer-continuous. A spacetime is said to be {\sl causally
continuous\/} if for all events $x \in M$, $\alex^+(x)$ and
$\alex^-(x)$ are outer continuous.  A globally hyperbolic spacetime is
causally continuous.

A convex normal neighbourhood (CNN) $\cnn \subset M$, is an open set
such that for any $p \in \cnn$, there exists a map $\exp_p$ from the
tangent space $T_p\cnn $ to $\cnn $ which is a diffeomorphism. An
important feature of $\cnn$ is that any two points in $\cnn$ are
joined by a unique geodesic which lies entirely in $\cnn$.

\noindent {\bf Some definitions from Riemannian geometry:} 

In a Riemannian space $(\Sigma, h)$, the distance function $d(x,y)$
between two points on $\Sigma$ is the length of the shortest path
between $x$ and $y$. A curve between $x,y$ is a {\sl segment} if its
length is $d(x,y)$. The {\sl convexity radius} at a $p \in \Sigma$ is
the largest $\cx_p$ such that the distance function $d(x,p)$ is a
convex function on the open ball $B(p,\cx_p)$ and any two points in
$B(p,\cx_p)$ are joined by unique segments lying entirely in
$B(p,\cx_p)$. The convexity radius $\cx$ of $\Sigma$ is given by the
infimum of $\cx_p$ for all $p \in \Sigma$.  Thus, if $d(p,q)< \cx$ for
any $p,q \in \Sigma$, then there exists a unique geodesic from $p$ to
$q$ of arc length $< \cx$.  An open set $ \cnn \subset \Sigma$ is said
to be {\sl convex with respect to $\cx$} if for any $p,q \in \cnn$,
the (unique) geodesic between them of arc-length $< \cx$ lies entirely
in $\cnn$. The intersection of two convex sets is then also convex
with respect to $\cx$.  A {\sl convex cover} of $\Sigma$ with respect
to $\cx$ is a locally finite cover of $\Sigma$ with each element a
convex set with respect to $\cx$.  $\diam(\cnn)$ is the {\sl diameter}
of an open set $\cnn$ which is the length of the longest geodesic
between any two points in $\cnn$ \cite{petersen}.

\section{The Nerve Simplicial Complex} 
\label{nerve}

We begin by constructing the nerve simplicial complex for an arbitrary
causal set.  Let $C$ be a causal set with $\can \subset C $ an \ia,
and $\cT_{\hv}(\can)$ an associated \ta{} for some ${\hv} \geq 0$ as
defined in (\ref{thickening}). Let $\cM$ be the set of maximal or
future-most elements in $\cT_{\hv}(\can)$ and $P_i \equiv
\ipast(m_i)\cap \ifut (A))$, $m_i \in \cM$.  The collection $\cP\equiv
\{P_i\}$ is a covering of $\cT_{\hv}(\can)$, i.e., $\bigcup_i P_i =
\cT_{\hv}(\can)$, since $P_i \subset \cT_{\hv}(\can)$ for all $i$ and
any $x \in \cT_{\hv}(\can)$ belongs to the inclusive past of some
maximal element of $\cT_{\hv}(\can)$.  For each $P_i$, define the
shadow sets $A_i \equiv P_i \cap A \subset \can$. Again, since $A_i
\subset \cT_{\hv}(\can)$ for all $i$, and any $a \in \can$ lies to the
inclusive past of a maximal element of $\cT_{\hv}(\can)$, the
collection $\bA \equiv \{ A_i \}$ covers $\can$.

The {\sl nerve} simplicial complex $\cN(\bA)$ of $\can$ is then
constructed by mapping each $A_i$ to a vertex, every
non-vanishing intersection $A_{i_0} \cap A_{i_1} \neq \emptyset$
to a 1-simplex, and in general, every non-vanishing intersection 
$A_{i_0} \cap A_{i_1} \ldots A_{i_k} \neq \emptyset$ to a
$k$-simplex \cite{rotman}. This construction is illustrated in
Fig. \ref{nerve.fig}.  The nerve simplicial complex
$\cN(P)$ of $\cT_{\hv}(\can)$ can be similarly constructed. 
\begin{figure}[ht]
\centering \resizebox{3.5in}{!}{\includegraphics{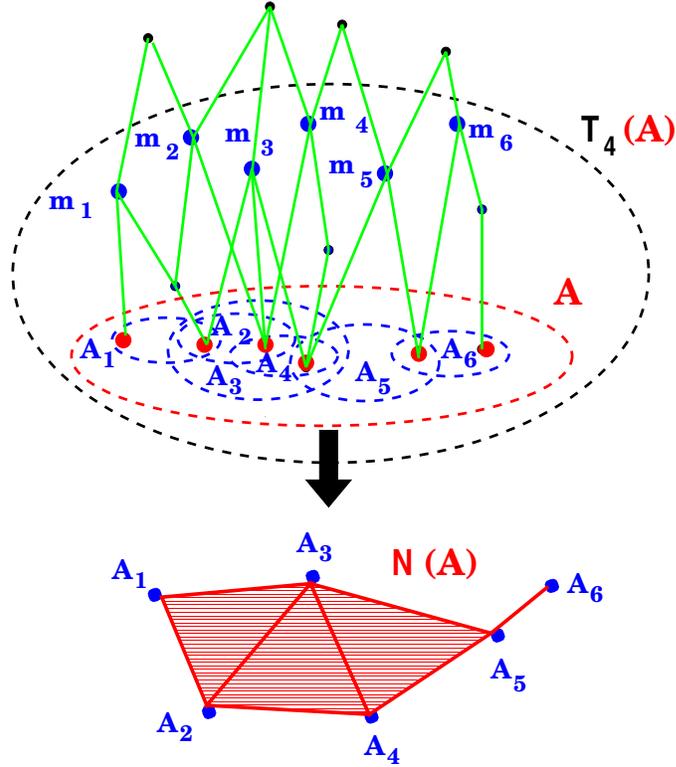}}
\vspace{0.5cm}
\caption{{\small Constructing a nerve simplicial complex $\cN(\bA)$
    from an inextendible antichain $\can$.}} \label{nerve.fig} 
\end{figure}

We now show that there exists a map $\Psi^\ast : \cN(\cP) \rightarrow
\cN(\bA)$ which is a bijection.  Let us define the projection map
$\Psi:\cP \rightarrow \bA $, i.e. $\Psi(P_i)=P_i \cap \can =A_i$. By
definition, $\Psi$ is onto. However, for an arbitrary causal set, it
is possible that $\Psi$ is not one-one: $P_i \cap \can = P_j \cap
\can$ need not imply that $i=j$, so that the shadows $A_i$ and $ A_j$
are identical as subsets of $\can$. Strictly speaking, then, the
collection $\bA=\{ A_i \}$ is a cover of $\can$, only if such
``redundant'' subsets are removed from it.  Let us however drop the
requirement that $\bA$ be a cover of $\can$, and retain these
redundant elements. Since now every $A_i$ comes from a unique $P_i$,
$\Psi$ is a bijection. Moreover, by set inclusion, any non-vanishing
intersection $A_{i_0\ldots i_k} \equiv A_{i_0} \cap A_{i_2} \ldots
\cap A_{i_k} \neq \emptyset$ has an associated non-vanishing
$P_{i_0\ldots i_k} \equiv P_{i_0} \cap P_{i_2} \ldots P_{i_k} \neq
\emptyset$. Hence, a $k$-simplex in $\cN(\bA)$ maps to a $k$-simplex
in $\cN(\cP)$, i.e., the map $\Psi^\ast: \cN(\bA) \rightarrow
\cN(\cP)$ is one-one. Moreover, if $P_{i_0\ldots i_k} \neq \emptyset$,
then $P_{i_0 \ldots i_k} \cap \can \neq \emptyset$: every $x \in
P_{i_0\ldots i_k}$ has a non-empty inclusive past in
$\cT_{\hv}(\can)$, $\ipast(x) \cap \ifut(\can)$, which, by
transitivity, is a subset of $P_{i_0\ldots i_k}$ which means that
$\exists \, a \in P_{i_0\ldots i_k} $ such that $a \in \can$. Or,
$P_{i_0\ldots i_k} \cap \can = (P_{i_0} \cap \can) \cap (P_{i_1} \cap
\can) \cap \ldots (P_{i_k} \cap \can)= A_{i_0 \ldots i_k} \neq
\emptyset $. Thus, a $k$-simplex in $\cN(\cP)$ maps to a $k$-simplex
in $\cN(\bA)$, which means that $(\Psi^\ast)^{-1}$ is one-one thus
establishing $\Psi^\ast$ as a bijection.

Preliminary numerical investigations for causets which faithfully
embed into 1+1  dimensional spacetimes show that $\cN(\bA)$ is
homologically equivalent to the continuum for a large range of values
of ${\hv}$ \cite{numerical}. In Fig \ref{graph.fig} we show the results
of numerical simulations for a causal set embedded into an $M= S^1
\times \re$ cylinder spacetime.  In the continuum, the non-vanishing
betti numbers are $b_0(M)= b_1(M)=1$, and there is no torsion. For $\hv
\leq 15$, $\cT_\hv(\can)$ splits into disconnected pieces, so that
$b_0(\cT) > b_0(M)$.  Similarly, for $\hv \geq 504$, $\{ b_i(\cT)\} \neq
\{ b_i(M) \}$. However, there exist a large range of thickenings,  $
16 \leq \hv \leq 503$, for which  $\{ b_i(\cT)\} = \{ b_i(M)\}$. 
\begin{figure}[ht]
\centering 
{\resizebox{6.0in}{!}{\includegraphics{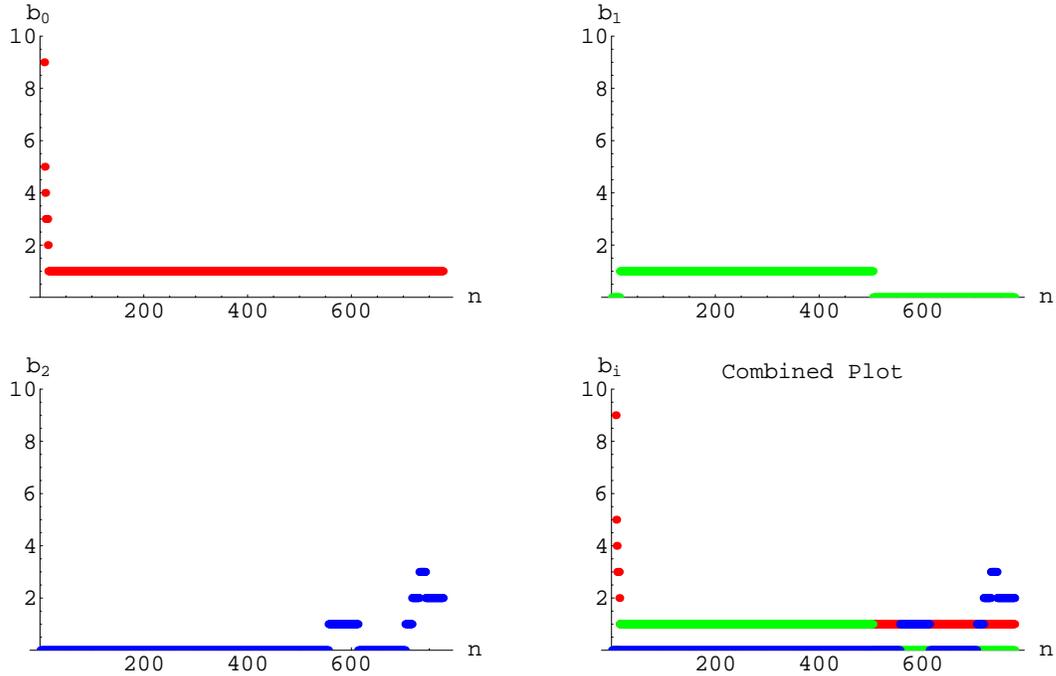}}}
\vspace{0.5cm} 
\caption{{\small The first three plots show the first three betti numbers $b_i$
    v/s thickening volume $\hv$ for a causet sprinkled onto a region
    of the cylinder  spacetime $S^1 \times \re$, whose non-vanishing
    betti numbers are $b_0= 1$, $b_1=1$. Superimposing these plots
    then helps us determine the range of thickenings ( $16 \leq \hv
    \leq 503 $) for which the \ta{} homology matches that of the
    continuum. (The torsion  vanishes uniformly and is hence not plotted.)  
 }}\label{graph.fig}
\end{figure}
Details will be discussed elsewhere \cite{numerical}, but for now, it
serves to justify our use of the discrete simplicial complex $\cN(\bA)$
as the appropriate starting point of our subsequent analytical investigations.

The first task is to find an appropriate continuum analog of the nerve
$\cN(\bA)$ and to ask how it is related to the spacetime topology.
For a globally hyperbolic spacetime $(M,g)$, we first note that a
Cauchy hypersurface $\Sigma$ is an appropriate continuum analog of the
\ia{} $\can$ since every point in $(M,g)$ is either to the future or
past of $\Sigma$ \cite{mots}. The continuum analog of the  \ta{}
$\cT_{\hv}(\can)$ can be obtained by correlating the number of
elements to the spacetime volume. Starting with a Cauchy hypersurface
$\Sigma$, one thus obtains a ``volume thickened'' region either to the
past or future of $\Sigma$. We now describe this construction in
detail.

Define the {\it volume} function $\vl$ on $M$ with respect to a  Cauchy
hypersurface $\Sigma$:  
\begin{equation} 
\vl(p)  =  \left\{ \begin{array}{ll} 
\vol(\alex(\Sigma,p)) & p \in \alex^+(\Sigma) \\  
- \vol(\alex(p,\Sigma)) & p \in \alex^-(\Sigma) \\ 
0  & p \in \Sigma, \qquad \end{array} \right. 
\label{zeta} 
\end{equation} 
where $\vol(X)$ denotes the spacetime volume of a region $X \subset
M$.  Let $\xi^a$ denote a continuous future-directed time-like vector
field on $(M,g)$ and $\xi(s)$ an associated integral curve of
$\xi^a$, where $s \in \Sigma$.  For a globally hyperbolic spacetime
$(M,g)$ with compact Cauchy surface $\Sigma$ we show that: \\

\begin{lemma} \label{continuity} $\vl(p)$ is a monotonically
increasing(decreasing) continuous function along the integral curves
$\xi(s)$ of $\xi^a$ to the causal future(past) of $\Sigma$, where $s
\in \Sigma$. 
\end{lemma} 
\bproof Monotonicity is immediate.  Let $\lambda$ be the parameter
along the integral curves $\xi_\lambda(s)$.  Let $q,r$ be two points
on $\xi_\lambda(s)$ which lie in $\alex^+(\Sigma)$, and such that $r
\in \alex^+(q)$. The interval $\alex(q,r)$ is non-empty and with
compact closure, since $(M,g)$ is globally hyperbolic. Hence
$\vol(\alex(q,r))$ is finite and non-zero.  Since $\alex(q,r) \subset
\alex(\Sigma, r)$, but $\alex(q,r) \nsubseteq \alex(\Sigma,q)$,
$\vl(r) \geq \vl(q) + \vol(\alex(q,r))$, or $\vl(r)>\vl(q)$. Since
there exists a $\lambda_1 < \lambda_2$ such that
$q=\xi_{\lambda_1}(s)$ and $r=\xi_{\lambda_2}(s)$, we see that $
\vl(\ld_1)\equiv\vl(q)< \vl(\ld_2)\equiv\vl(r) $ for any $\lambda_1 <
\lambda_2$.

Next, we show that $\vl$ is a continuous function on $\xi(s)$.  For a
{\it given} $s \in \Sigma$, the points in $\xi(s)$ can simply be labelled by
the parameter $\lambda$. Let $\lambda_0 $ such that 
$\xi_{\lambda_0}(s)$ lies in $\alex^+(s)$.  
Define $\bar \alex_\pm \equiv \limit_{\epsilon \rightarrow 0
}\alex(\Sigma,\ld_0\pm \epsilon)$ for $\epsilon>0$ and $\vl_\pm \equiv
\vol(\bar \alex_\pm)= \vol(\limit_{\epsilon \rightarrow 0}
\alex(\Sigma, \lambda_0 \pm \epsilon)) = \limit_{\epsilon \rightarrow
0} \vl (\ld_0 \pm \epsilon)$.
Monotonicity implies that $\vl(\ld_0 -\epsilon) < \vl (\ld_0) <
\vl(\ld_0+\epsilon) $ for $\epsilon>0$ and hence $\vl_- \leq
\vl(\lambda_0) \leq \vl_+$. 

Let $\vl_- < \vl(\lambda_0)$ be a strict inequality. 
This means that  $\alex(\Sigma, \ld_0)$ contains a set $S$ of finite 
volume such that $S \nsubseteq \alex(\Sigma, \lambda_0 -
\epsilon)$ for {\it any } $\epsilon >0$. For any compact subset $K
\subset S$, this implies that  
$K \subset \alex^-(\ld_0)$, but $K \nsubseteq \alex(\Sigma, \lambda_0 -
\epsilon)$ $\forall \,\, \epsilon >0$.
Using the set-decomposition $\alex^-(\lambda_0 -\epsilon) =
\alex(\Sigma, \lambda_0 -\epsilon) \sqcup (\alex^-(\lambda_0
-\epsilon) \cap \Sigma) \sqcup \alex^-((\alex^-(\lambda_0-\epsilon)
\cap \Sigma))$, we see that since $K$ neither is in or to the past of
$\Sigma$, $K \nsubseteq \alex^-(\lambda_0 - \epsilon)$. Thus, even
though $K \subset \alex^-(\lambda_0)$, every neighbourhood  of
$\ld_0$ contains a $p \in \alex^-(\lambda_0)$ such that $K \nsubseteq
\alex^-(p)$, i.e., $\alex^-(\lambda_0)$ is not inner continuous, which
is a contradiction \cite{HS}. 

Now let $\vl_+ > \vl(\lambda_0)$ be a strict inequality. This means
that $\alex_+$ is not simply the closure of $\alex^-(\Sigma, \ld_0)$,
and hence there exists an $S \subset \inte(\alex_+) $ of finite volume
such that $ S \nsubseteq \overline{\alex(\Sigma, \ld_0)}$.  Since
$\overline{\alex^-(\lambda_0)} = \overline {\alex(\Sigma, \lambda_0)}
\cup \overline{\alex^-(\ld_0) \cap \alex^-(\Sigma)}$ and $S$ is
neither in nor to the past of $\Sigma$, $S \nsubseteq
\overline{\alex^-(\lambda_0)}$.  Thus, for any compact $K \subset S$,
$K \nsubseteq \overline{\alex^-(\lambda_0)}$. However, $K \subset
\alex(\Sigma, \lambda_0+ \epsilon)$ for all $\epsilon >0$, which means
that every neighbourhood of $\lambda_0$ contains a $p \in
\alex^+(\ld_0)$ such that $K \subset \alex^-(p)$, which means that
$\alex^-(\ld_0)$ is not outer continuous. For $(M,g)$ globally
hyperbolic, this is not possible.

Thus,  $\vl_+ = \vl(\lambda_0) =\vl_-$, and hence $\vl(\lambda)$ is
a continuous function. The proof for the time-reversed case is
identical. 
\eproof

Since $\vl$ is a continuous, monotonically increasing function along
$\xi(s)$ to the future of $\Sigma$, we can reparameterise the curves
$\xi_\lambda(s)$ to $\xi_\vl(s)$. $\xi_\vl(s)$ then provides us with a
one-parameter family of homeomorpisms between $\Sigma$ and $\vl
=\mathrm{const}$ hypersurfaces $\Sigma_\vl$ to its future.
The spacetime region $M_\vl \equiv \overline{\alex(\Sigma,
\Sigma_\vl)}$ can be thought of a volume thickening of $\Sigma$ and is
the desired continuum analog of a thickened antichain.

The continuum nerve simplicial complex analogous to $\cN(\bA)$ may
then be constructed from $M_\vl$ as follows. $\Sigma_\vl$ are the
maximal or future most elements of $M_\vl$, and the past of any point
$p \in \Sigma_\vl$ casts a {\sl shadow} 
\begin{equation}\label{shadow}  
\sigma(p)=\alex^-(p) \cap \Sigma
\end{equation} 
on $\Sigma$. The set $\mS_\infty \equiv \{ \sigma(p) | p \in
\Sigma_\vl\}$ is an open cover of $\Sigma$, since for every $s \in
\Sigma $ there is a $p\in \Sigma_\vl$ such that $ s \in I^-(p)$. The
associated nerve $\cN(\mS_\infty)$ is an infinite dimensional
simplicial complex, arising from non-vanishing intersections of
arbitrarily many sets in $\mS_\infty$.

In the continuum limit in which the discreteness scale is taken to
zero, $\cN(\bA) \rightarrow \cN(\mS_\infty)$.  However, since our
interest lies in the continuum {\it approximation} for which the
discreteness scale is not taken to zero, it is more appropriate to use
a locally finite subcover of $\mS_\infty$.  We do so by choosing a
collection of points $\{m_i\}$ on $\Sigma_\vl$ such that
(a) the collection of shadows $\mS \equiv \{ \sigma(m_i)\}$ is a cover of
$\Sigma$ and (b) Any open set in $\Sigma_\vl$ contains only a finite
number of $m_i$.  Then $\mS$ is a locally finite cover of
$\Sigma$, and $\cN(\mS)$ is the required continuum analog of
$\cN(\bA)$.

\section{A Continuum Result} 

Of course, one of the main questions we want to address is whether
such a continuum nerve construction yields a simplicial complex which
is homological to $\Sigma$ and hence to $M$. For example, in order to
construct the C\v{e}ch cohomology, one starts with an ordered infinite
collection $\{\mS_1, \mS_2 \ldots \mS_i \ldots \} $ of locally finite
open coverings of $M$, such that $\mS_{i+1}$ is a refinement of
$\mS_i$. The correct continuum cohomology is then recovered only in
the limit $i \rightarrow \infty$, i.e. the limit of infinite
refinement. This is reasonable in the continuum, since any
finite refinement can contain sets that are ``too large'', with the
result that the nerve simplicial complex masks non-trivial topological
information.

However, for any fundamentally discrete theory like causal set theory,
continuum structures smaller than the discreteness scale are
unphysical. Hence, it should suffice to consider a finite cover. In
particular, e wish to ask what topological information of $\Sigma$ is
encoded in the nerve $\cN(\mS)$ of the locally finite cover $\mS
\equiv \{ \sigma(m_i)\}$ of $\Sigma$. More specifically, is $\cN(\mS)$
homologous to $\Sigma$.  A look at a simple example tells us that the
answer in general is no.

Consider the $S^1 \times \re $ cylinder spacetime $ds^2=-dt^2+
d\theta^2$, with $\theta \in [0, 2 \pi]$. If $\Sigma$ is the $t=0$
circle, then $\vl= \pm t^2 $ if $t < \pi$. For $t= 3 \pi/4$, let us
construct a cover $\mS_1$ as follows. The shadows from the pair of
points $m_1=(3 \pi/4,0)$ and $m_2=(3 \pi/4 , \pi)$ are sufficient to
cover the $t=0$ slice and $\sigma(m_1) \cap \sigma(m_2)$ is the
disjoint union of the two open intervals $(\pi/4, 3 \pi/4) \sqcup (5
\pi/4, 7 \pi/4)$ on the $t=0$ circle. $\cN(\mS_1)$ is then a single
1-simplex, which is not homological to $S^1$. On the other hand, for
$t= \pi/2$, a cover $\mS_2$ is provided by the shadows from the three
points $m_1=(\pi/2,0), m_2=(\pi/2,2\pi/3), m_3=(\pi/2,4\pi/3)$. Their
intersections on the $t=0$ slice are the intervals, $\sigma(m_1)
\bigcap \sigma(m_2) = (\pi/6,\pi/2)$, $\sigma(m_2) \bigcap \sigma(m_3)
= (5 \pi/6, 7 \pi/6)$ and $\sigma(m_3) \bigcap \sigma(m_1)=(3 \pi/2,
11 \pi/6)$, respectively, from which we see that
$\bigcap_{i=1}^3\sigma(m_i) = \emptyset$. $\cN(\mS_2)$ is then the
boundary of a $2$-simplex and therefore homological to $S^1$. This
simple example shows us that the choice of cover $\mS$ is crucial for
determining whether $\cN(\mS)$ has the correct homology.

A somewhat lesser known result in algebraic topology due to DeRham and
Weil \cite{dederham} then gives us a criterion for $\mS$ such that
$\cN(\mS)$ is not only homological, but homotopic to $\Sigma$.
 
\begin{theorem}\rm{{\bf {(De Rham-Weil)}}}  \label{dederham} 
The nerve of a convex cover of $\Sigma$ is homotopic to
$\Sigma$. 
\end{theorem}

Theorem \ref{dederham} is also valid for a {\sl simple cover}, whose
elements are differentiably contractible as are all intersections
\footnote{Since our interest here is restricted to homology, it
suffices that the intersections are acyclic, i.e. they contain no
non-trivial cycles.}. From this we see why
the two covers on the cylinder spacetime give such different results.
While the shadows are themselves differentiably contractible in the
cover $\mS_1$, their intersections are homotopic to two disconnected
points, so that $\mS_1$ is neither convex nor simple. For $\mS_2$, the
shadows and their intersections are both convex and contractible,
resulting in a nerve which is homological to $\mbs^1$ in accordance with
Theorem \ref{dederham}.

The open cover of interest to us is the collection of shadow sets
$\mS$. As we have seen in the cylinder spacetime, this is is not
always a convex cover. Nevertheless, as is apparent even for this
simple spacetime, for times $t \leq \pi/2$, every shadow is in fact a
convex set of the $t=0$ slice. This suggests that for $\vl$ below some
critical value, one should always get shadows that are convex, for any
globally hyperbolic spacetime. We will now show that this is indeed
the case when the Cauchy surface $\Sigma$ is taken to be compact. We
will take $(M,g)$ and $(\Sigma,h)$ to be at least $C^2$
differentiable.

Let $\cx$ be the convexity radius on the Cauchy slice, $(\Sigma, h)$,
where $h$ is the induced spatial metric.  We again take $\xi^a$ to be
a continuous, future-directed timelike vector field and $\xi(s)$ its
integral curve which intersects $\Sigma$ at $s$. We use the volume
function $\vl$ as defined in Eq (\ref{zeta}) to parameterise this
curve and denote the shadows of the past sets of $\xi_\vl(s)$ on
$\Sigma$ by $\sigma_s(\vl)$ for $\vl > 0$.  We also note that for
$\Sigma$ compact, there exists a $\vcrit$ such that $\forall \, \, \vl
\geq \vcrit$, there exist $p \in \Sigma_\vl$ such that
$\sigma(p)=\Sigma$.

Now, any $s \in \Sigma$ is contained in a CNN $\cnn$ of $(M,g)$ such
that $\cnn \cap \Sigma$ is also a CNN of $(\Sigma,h)$. We will term
such an $\cnn$ a {\sl Cauchy-CNN} or C-CNN. One way of constructing
such a C-CNN is from a sufficiently small CNN $\cnnn \ni s$ of
$(\Sigma, h)$, such that its domain
of dependence $D(\cnnn)$  is itself a  CNN of $(M,g)$.

\begin{lemma}\label{cnn}  
For every $s \in \Sigma$ and $\xi^a$ a future-directed time-like
vector field, $\exists$ a $\vl_s$, $\vcrit> \vl_s > 0$, such that for
any $0 < \vl < \vl_s$,
\begin{enumerate}
\item 
  $\diam(\overline{\sigma_s(\vl)}) < \cx$. 
\label{nesting}  
\item $\pdsz$ has positive principal
  curvatures with respect to its outward normal, for spacetime
  dimension $\dimm \geq 3$.   \label{principal}
\end{enumerate} 
\end{lemma} 
\bproof We first show that for any $\vl_1,\vl_2$, $\vcrit> \vl_1>
\vl_2 > 0$, $\sigma_s(\vl_2) \subset \sigma_s(\vl_1)$.  By
transitivity, $ \overline{\sigma_s(\vl_2)} \subseteq
\overline{\sigma_s(\vl_1)}$. The strict inclusion is proved as
follows. Assume that $\sigma_s(\vl_2)=\sigma_s(\vl_1)$ for $\vl_2 <
\vl_1$. Any $x \in \pdszt$ lies on $\pd \overline{(\alex^-(\vl_2))}$
which by global hyperbolicity means that $x \rightarrow
\xi_{\vl_2}(s)$, i.e., they are null-related. Since $\xi_{\vl_2}(s)
\prec \prec \xi_{\vl_1}(s)$, this means that $x \prec \prec
\xi_{\vl_1}(s)$ or $x$ lies in $\alex^-(\xi_{\vl_1}(s))$
\cite{HE}. But $x \in \pd \overline{(\sigma_s(\vl_2))} \subset \pd
\overline{(\alex^-(\vl_1))}$ as well, which is a contradiction.
Therefore, $\overline{\pd \sigma_s(\vl_2)} \cap \overline{\pd
\sigma_s(\vl_1)} = \emptyset$ and hence $\overline{\sigma_s(\vl_2)}
\subset \overline{\sigma_s(\vl_1)}$, which implies that
${\sigma_s(\vl_2)} \subset {\sigma_s(\vl_1)}$, since $\sigma_s(\vl)$
is an open set. Thus, the $\sigma_s(\vl_i)$ are nested one inside the
other.

Now, for every CNN $\nb' \ni s$, in $(\Sigma, h)$, there exists a
$\vl_s' >0$ such that $\sigma_s(\vl) \subset \nb'$: Let
$D^+(\overline{\nb'})$ be the future domain of dependence of
$\overline{\nb'}$ (a closed achronal set), and $H^+(\overline{\nb'})$
its future boundary.  Let $p \in \xi_\vl(s)$ such that $s\prec
\prec p \prec \prec r$, where $r \in \xi_\vl(s) \cap
H^+(\overline{\nb'}) \neq \emptyset$. Then, $\alex^-(p) \cap \Sigma=
\sigma_s(\vl(p))\subset \overline \nb'$ since $p \in
D^+(\overline{\nb'})$, so that $\overline{\sigma_s(\vl(p))}\subseteq
\overline \nb'$.  Assume equality. Then for all $x \in \pd
\overline{\sigma_s(\vl(p))} = \pd \overline{\nb'}$, $x\rightarrow p
\prec \prec r$, or $x \prec \prec r$. But since $r \in
H^+(\overline{\nb'})$, by globally hyperbolicity, there exists an $x'
\in \pd{\overline{\nb'}}$ for which $x' \rightarrow r$, which implies
a contradiction. Hence $\sigma_s(\vl(p))\subset \overline \nb'$.
Thus, for all $\vl < \vl_s' \equiv \vl(r)$, $\sigma_s(\vl) \subset \nb'$.

Moreover, let $\nb'$ such that $\diam(\overline{\cnn'}) < \cx$. Again,
$\vl_s' >0 $ is such that for all $0< \vl < \vl_s'$, $\sigma_s(\vl)
\subset \cnn'$. Let $p, q \in \pd \overline{ \sigma_s(\vl)}$ such that
$d(p,q)=\diam(\overline{\sigma_s(\vl)})$ which is the length of the
(unique) geodesic $\gamma$ in $\cnn'$ from $p$ to $q$. Let
$\gamma(0)\equiv p$, $\gamma(1)\equiv q$ and $\gamma(1+\epsilon)
\equiv r$. For small enough $\epsilon>0$, $r \in \cnn'$. $d(p,r)$ is
therefore the arc-length of $\gamma$ from $p$ to $r$, so that $d(p,r)
= d(p,q) + d(q,r)$, thus implying that $\cx > \diam(\overline{\cnn'})
\geq d(p,r) > \diam(\overline{\sigma_s(\vl)})$. This proves point
\ref{nesting}.

We can do better, i.e., find the largest possible $\vl_s'$ for which
point \ref{nesting} is true by taking the supremum over all such
C-CNN's $\cnn'$, i.e., $\vl_s'^{\sup} \equiv \sup_{\cnn'} \vl_s'>0$. 

To prove point \ref{principal} we note that for a flat (i.e. zero
extrinsic curvature) Cauchy hypersurface in Minkowski spacetime, the
principal curvatures of the boundary of the shadows of a past light
cone on it, is strictly positive. We show that we can construct
``small enough'' neighbourhoods of $s$ such that the deviation from
flatness is sufficiently small for this to possible.

Since the boundaries of the shadows $\pd \overline{(\sigma_s(\vl))}$
do not intersect for different $\vl<\vcrit$ by outer continuity, the
$n-2$ dimensional surfaces $\pd \overline {\sigma_s(\vl)} \simeq
\mbs^{n-2}$ provide a foliation of $\Sigma$ centered at $s$.  The
boundary of $\jj^-(\xi_\vl(s))$ is generated by past directed null
geodesics which intersect $\Sigma$ at $\pd
\overline{\sigma_s(\vl)}$. Let ${\vec \sy}$ coordinatise these $n-2$
null-directions, i.e., the $\mbs^{n-2}$.  In $2+1$-dimensions, for
example, ${\vec \sy}=\theta$ the coordinate on a circle, $\mbs^1$.
Let $\fE$ be the set of monotonically increasing functions of $\vl$
which vanish at $\vl =0$. Since the deviation from flatness decreases
with $\vl$, a given function in $\fE$ can serve to parametrise the
deviation from flatness of the metric in a CNN of $s$ in $(\Sigma,h)$
along any given null direction $\vec \sy$ projected onto $\Sigma$.

The set $(\vl, {\vec \sy})$ are therefore Riemman normal coordinates
on a CCN $\cnn'$ of $(\Sigma, h)$, centered at $s$ and the induced
spatial metric is $h_{ab}(\vl, {\vec \sy}) = \delta_{ab} + \cO(\ef)$,
$\ef(\vl) \in \fE$, where the ${\vec \sy}$ dependence on the right
hand side is absorbed into the second term. Let $n^a$ be the unit
normal to $\cnn'$ in $(M,g)$, so that the extrinsic curvature of
$\cnn'$ is $K_{ab} = \frac{1}{2}{\cL}_{n}h_{ab}$.  Choose $\cnn'$ to
be small enough so that $n^{a}$ is nearly constant over $\cnn' $, i.e.
$n^{a}(\vl,{\vec \sy}) = n^{a}(s) + \cO(\ef')$ with $\ef'(\vl) \in
\fE$, again with the ${\vec \sy}$ dependence absorbed into the second
term. Since $\Sigma$ is smoothly embedded into $(M,g)$, $n^{a}$ and
hence $K_{ab}$ vary smoothly in $(M,g)$. The Christoffels
$\Gamma^{c}_{ab}(\vl,{\vec \sy})= \cO(\ef)$ and vanish at $s$, so that
extrinsic curvature $K_{ab}= \cO(\tef)$, where $\tef$ is the more
dominating of the two functions $\ef$, $\ef'$, as $\vl \rightarrow 0$.

If $l_{ab}(\vl,{\vec \sy})$ is the induced metric on $\pdsz$ and
$m^a(\vl)$ its outward normal, the foliation by $\pdsz$ allows us to
express $h_{ab}(\vl)= l_{ab}(\vl,{\vec \sy}) + m_a(\vl,{\vec
\sy})m_b(\vl,{\vec \sy})$.  This means that $l_{ab}(\vl,{\vec \sy})$
and $m^a(\vl,{\vec \sy})$ each differ from their flat space
counterparts, i.e., on $(\cnn', \delta)$, by at most some
$\cO(\ef_m)$, $\ef_m \in \fE$, where $\ef_m$ can differ from $\ef$ if
it is more dominating as $\vl \rightarrow 0$. Thus, $l_{ab}(\vl, {\vec
\sy}) =l_{ab}^\mF(\vl, {\vec \sy})+ \cO(\ef_m)$ and $m_a(\vl, {\vec
\sy}) = m_a^\mF(\vl, {\vec \sy})+ \cO(\ef_m)$, where the $\mF$ labels
the quantities with respect to the flat metric $\delta_{ab}$ on
$\cnn'$.  The principle curvatures of $\pdsz$ are therefore
$\pc_\ind(\vl,{\vec \sy}) = \pcf_\ind(\vl,{\vec \sy}) + \cO(\ef_m)$,
$\ind=1, \ldots n-2$, with $\pcf_\ind(\vl,{\vec \sy})$ the principle
curvature of these surfaces with respect to $\delta_{ab}$ on $\cnn'$.
For example, for a spherically symmetric foliation about $s$
in a flat metric, the $\pcf_\ind(\vl)$ are positive and diverge to $+
\infty$ as $\vl \rightarrow 0$. 

In calculating the dominant contributions to the principal curvatures
$\pc_\ind$ we have not yet made any assumptions about the foliation $\{ \pd
\overline{(\sigma_s(\vl))}\}$ which depends on the spacetime metric.
Without this further input, the $\pcf_\ind$ will not in general
be positive.

Let us therefore choose a C-CNN $\cnn \ni s$ in $(M,g)$ such that not
only does $\cnn' \equiv \cnn \cap \Sigma$ satisfy the above
conditions, but the spacetime metric $g_{ab} = \eta_{ab} + \cO(\ef'')$
in $\cnn$ with $\ef''(\vl) \in \fE$.  For example, for $\cnn$ strictly
Minkowskian, and $\cnn'$ with zero extrinsic curvature, the $\pd
\overline{\sigma (\vl_F)}$ have strictly positive $\pcf_\ind$, where
$\vl_F$ represents is the equivalent volume in Minkowski
spacetime. Thus, the volume $\vl= \vl_F + \cO({\ef''})$, and the
principle curvatures of $\pdsz$, $\pc_\ind(\vl, {\vec \sy}) =
\pcf_\ind(\vl_\mF, {\vec \sy}) + \cO({\tef})$, where $\tef$ is now the
more dominating of the functions $\ef_m, \ef', \ef''$ as $\vl
\rightarrow 0$.  We therefore see that for any fixed null direction
${\vec \sy}$ there exists a $\vl({\vec \sy}) >0$ such that for all $0<
\vl< \vl({\vec \sy})$, $\pc_\ind(\vl,y^a) >0$. Since $\Sigma$ is a
smooth Cauchy hypersurface, the boundaries of the shadows are smooth
in a C-CNN. Thus, $\pc_\ind(\vl, \vec \sy)$ is continuous with respect
to $\vec \sy$, i.e., over the set of null directions $\mbs^{n-2}$. If
$\vl < \vl(\vec \sy) $ for some $\vec \sy \in \mbs^{n-2}$, then since
$\pc_\ind(\vl, \vec \sy) > 0 $ there exists a neighbourhood $U$ of
$\vec \sy$ in $\mbs^{n-2}$ such that for all $\vec \sy' \in U$,
$\pc_\ind(\vl, \vec \sy')>0 $. Thus, $\vl(\vec \sy') \geq \vl(\vec
\sy)$.  Now, for every $\vec \sy$, $\vl(\vec \sy)>0 $.  However, if
$\inf_{\vec \sy} \vl(\vec \sy) =0$, then there is an infinite sequence
$\{ \vl(\vec \sy_i)\} \rightarrow 0$. The corresponding sequence in
$\mbs^{n-2}$, $\{\vec \sy_i \} \rightarrow \vec \sy_0 \in
\mbs^{n-2}$. Thus, for every neighbourhood $U$ of $\vec \sy_0$ there
exists a $j$ such that for all $i>j$, $\vl(\vec \sy_i) < \vl(\vec
\sy_0) $ and $\vl(\vec \sy_0) > 0$. This is a contradiction. Hence,
$\vl_s''=\inf_{\vec \sy}\vl(\vec \sy)>0$, 
which proves point \ref{principal}.

We may now obtain the largest possible $\vl_s''$ by now varying over
all $\cnn'$ which satisfy the above criteria, to obtain
$\vl_s''^{\sup}=\sup_{\cnn'} \vl_s''>0$.

Thus, for every $s\in \Sigma$, there exists a $\vl_s =
\min(\vl_s'^{\sup}, \vl_s''^{\sup}) >0 $ for which \ref{nesting} and
\ref{principal} is satisfied.  \eproof

For a fixed time-like vector field $\xi^a$, although $\vl_s>0$ for
each $s \in \Sigma$, we now show that $\inf_{s} \vl_s > 0$ using
global hyperbolicity and the compactness of the Cauchy hypersurfaces
$\Sigma$. Let $\gamma(\alpha)$ be a continuous curve in $(\Sigma,h)$
through $s \in \Sigma$, with $\gamma(0)=s$. The homemorphism
$\xi_v:\Sigma \rightarrow \Sigma_v$ takes $\gamma(\alpha) \rightarrow
\gamma_v(\alpha)\equiv \xi_v \circ \gamma (\alpha)$. Let
$S(\alpha)\equiv \overline{\sigma_{\gamma(\alpha)}(v)}$, i.e., the
closure of the shadows of past light cones of points in
$\gamma_v(\alpha)$ onto $\Sigma$. $\{ \pd S(\alpha) \} $ is then a
family of $\mbs^{n-2}$'s. We now prove a couple of claims.
\begin{claim}
The family $\{ S(\alpha) \}$ is continuous with respect to $\alpha$. 
\end{claim} 
\bproof Let $f(\alpha, \vl, \vec \sy)=0$ be the functions defining the
$\pd S(\alpha)$'s.  Assume that $f$ is not continuous at $s$ or
$\alpha=0$, so that $\lim_{\alpha \rightarrow 0} f(\alpha, \vl, \vec
\sy) = \tf(0, \vl, \vec \sy) \neq f(0, \vl, \vec \sy)$. Equivalently,
$\lim_{\alpha \rightarrow 0} S(\alpha) = \tS(0) \neq S(0)$. Let
$\{s_i\}$ be an infinite sequence of points on $\gamma(\alpha)$ which
converge to $s$ with $s_i \equiv \gamma(\alpha_i)$ and $0<
\alpha_{i+1} < \alpha_i$. By definition $\tS(0)$ is the limit set of
$\{ S(\alpha_i)\} $, i.e., for every $q \in \tS(0)$ and any open $O
\ni q$ there exists a $j$ such that forall $i>j$, $O \cap S(\alpha_i)
\neq \emptyset$. Let $\{p_i\}$ be the corresponding sequence on
$\gamma_\vl(\alpha)$, i.e., $p_i \equiv \xi_\vl(s_i)$, which converges
to $p \equiv \gamma_v(0)$.

Now, let $q \in \tS(0)\backslash S(0)$, which is non-empty by
definition. Since $S(0)$ is closed, there exists an open $O \ni q$
such that $O \cap S(0)=\emptyset$. Any compact subset $K$ of $O$
therefore lies in $S(0)^c$, and hence in $\overline{\alex^-(p)}^c
$. Let $q \in K \subset O$ and let $q \in O' \subset K$ be open. For
any such $q$, $O'$, then, there exists a $j$ such that $\forall \, \,
i>j$, $O' \cap S(\alpha_i) \neq \emptyset$ since $q \in
\tS(0)$. Therefore $K \cap S(\alpha_i) \neq \emptyset$ and $K \cap
\overline{\alex^-(p_i)} \neq \emptyset$. Thus, every neighbourhood of
$p$ contains a $p_i$ such that $K \cap \overline{\alex^-(p_i)}^c \neq
\emptyset$.  This contradicts causal continuity.  
\eproof

\begin{claim} 
If $\vl < \vl_s$ then $\exists \, \, \delta > 0$ such that $\forall \,
\, 0 < \alpha < \delta$,  $S(\alpha)$ satisfies point \ref{nesting} and
 \ref{principal} of Lemma \ref{cnn}.  
\end{claim}    
\bproof Continuity of the $S(\alpha)$ with respect to $\alpha$ means
continuity of its associated geometric properties. Since $S(\alpha)$
are continuous, there exists a homemorphism $\varphi_{\alpha}: S(0)
\rightarrow S(\alpha)$. Let $\chi(0,t)$ be a (continuous) segment from
$x(0)$ to $y(0)$ on $S(0)$, mapped by $\varphi_{\alpha} $ to $
\chi(\alpha, t)$ with end points $x(\alpha),y(\alpha) \in
S(\alpha)$. The length $\ell(\chi(\alpha,t))$ is itself a continuous
function of $\alpha$ because of the continuity of the metric $h_{ab}$
on $\Sigma$.  Hence, since $\ell(\chi(0,t)) < \cx$, this means that
there exists a $\delta>0$ such that for all $0< \alpha < \delta$,
$\ell(\chi(\alpha,t)) < \cx$. Therefore $d(x(\alpha), y(\alpha)) <
\cx$.  Again, continuity of $h_{ab}$ and the $\pd S(\alpha)$ means
that the principal curvatures $\pc_\ind(\alpha)$ are continuous with
respect to $\alpha$.  Since $\pc_\ind(0)>0$, there exists a
$\delta'>0$ such that $\forall 0< \alpha < \delta'$,
$\pc_\ind(\alpha)>0$ for all $i$.

Thus, there exists a $\delta''=\min(\delta, \delta')>0$ such that
$\forall \, \, 0< \alpha < \delta''$ point \ref{nesting} and
\ref{principal} of Lemma \ref{cnn} are satisfied.  \eproof

Thus, for every $s \in \Sigma$, there exists a neighbourhood $O$ of
$s$ such that for all $s' \in O$, $\vl_s \leq \vl_{s'}$. Since
$\Sigma$ is compact, an argument similar to the one used to show
$\vl_s''>0$ in the proof of Lemma \ref{cnn} can be used to show that
$\tvl'\equiv \inf_{s \in \Sigma} \vl_s > 0$.

By varying over the time-like vector fields $\xi^a$, we obtain a
{\sl convexity volume} of $\Sigma$ which is the largest $\vl'$ for a 
given $(\Sigma,h) \subset (M,g)$
\begin{equation} 
\tvl \equiv \sup_{\xi^a} \tvl', \quad  0 < \tvl \leq  \vcrit.  
\end{equation}  

The positivity of the principle curvatures of the boundaries of the
shadows, then implies, by the maximum principle \cite{petersen}, that
for any $p,q \in \sigma_s(\vl)$ and $0< \vl < \tvl$, if there exists
a geodesic from $p$ to $q$ of arc-length $< \diam(\sigma_s(\vl))$,
then it is unique and lies in $\sigma_s(\vl)$ for $\dimm \geq 3$. For $\dimm
=2$, we don't require \ref{principal} since as long as \ref{nesting}
is satisfied, any geodesic from $p$ to $q$ in $\sigma_s(\vl)$, of
arc-length $< \diam(\sigma_s(\vl))$ must lie in $\sigma_s(\vl)$. 
Hence, $\sigma_s(\vl)$ is convex with respect to $\cx$ for all $\dimm
\geq 2$.

Thus, the collection $ \mS^\infty_\vl \equiv \{\sigma_s(\vl)\} $, for
any $\vl < \tvl$ forms an open covering of $\Sigma$ whose elements are
convex sets with respect to $\cx$. A locally finite cover $\mS_\vl$
can be obtained from $\mS^\infty_\vl$ as follows. Since $\Sigma$ is
paracompact, $\mS^\infty_\vl $ admits a locally finite refinement,
i.e. a locally finite cover $\caV$ of $\Sigma$ such that for every
$V_i \in \caV$, there exists a $\sigma_s(\vl) \in \mS^\infty_\vl$ such
that $V_i \subset \sigma_s(\vl)$. For every $V_i$, choose an $s_i$
such that $V_j \subset \sigma_{s_i}(\vl)$ implies $j=i$. Since $\caV$
is a cover of $\Sigma$, so is $\mS_\vl \equiv \{
\sigma_{s_i}(\vl)\}$. Local finiteness of $\caV$ implies that every $p
\in \Sigma$ has a neighbourhood $\cnn \ni p$ which intersects only a
finite number of $V_i$. Since each $V_i$ is contained in a unique
$\sigma_{s_i}(\vl)$, $\cnn$ also intersects only a finite number of
$\sigma_{s_i}(\vl)$, so that $\mS_\vl$ is a locally finite convex
cover of $\Sigma$.  $\mS_\vl $ is therefore a convex open cover of
$\Sigma$ and hence, using Theorem \ref{dederham} we have shown that
\begin{theorem}\label{main} The nerve $\cN(\mS_\vl)$ is homotopic to
  $\Sigma$ and thence to $M$.
\end{theorem} 

In order to make the continuum-discrete comparison, which is the main
goal of this paper, one would like to obtain the discrete analog of
the cover $\mS_\vl$. While it is tempting to use the obvious
identification with the collection $\bA$ as suggested at the end of
Section \ref{nerve}, one must proceed with caution. Since the
continuum-discrete correspondence for causets maps {\it spacetime}
volume (and {\it not} spatial volume) to cardinality, we need to find
a collection of open sets in spacetime corresponding to $\mS_\vl$. The
natural choice is the collection of past sets, $ \bI_\vl \equiv \{
\alex_i \equiv \alex(\Sigma, \xi_\vl(s_i))\}$.  For ease of notation,
let $\xi_i\equiv \xi(s_i)$, $m_i =\xi_\vl(s_i)$ and $\sigma_i\equiv
\sigma_\vl(s_i)$.  Since $\sigma_i =\alex^-(m_i) \cap \Sigma$, the map
$\Psi: \bI_\vl \rightarrow \mS_\vl$ is onto. If $\Psi$ were in
addition one to one, then $\sigma_i = \sigma_j \Rightarrow i=j$. Assume
otherwise, i.e.  $\sigma_i = \sigma_j, \, \, i\neq j $, so that $ \pd
\overline{\sigma_i}= \pd \overline{\sigma_j}$.  Since the spacetime is
globally hyperbolic $\forall \,\,\, x \in \pd \overline{\sigma_i}$, $x
\rightarrow m_i$ and $x \rightarrow m_j$. Now, $s_i, s_j \in
\alex^-(m_i)\cap \alex^-(m_j)$,  which means that $\xi_j \cap
(\alex(\Sigma,m_i) \cap \alex(\Sigma,m_j)) \neq \emptyset$.  Since
$m_i$ and $m_j$ are spacelike to each other, $\xi_j$ must  
leave $\pd \overline{\alex(\Sigma,m_i)}$ at some $y = \xi_j \cap \pd
\overline{ \alex^-(m_i)}$.  Therefore, $y \rightarrow m_i$ and $ y
\prec \prec m_j$. Since $y \in \pd \overline{\alex^-(m_i)}$, there
exists $x \in \pd \overline{\sigma_i}$ such that $x \rightarrow y
\rightarrow m_i$. However, since $x \rightarrow y \prec \prec m_j$,
this means that $ x \prec \prec m_j$ \cite{HE}, which is a
contradiction. Hence $\Psi$ is one-one.

Consider the non-empty intersection $ \sigma_{i_1 \ldots i_k} \equiv
\sigma_{i_1} \bigcap \ldots \bigcap \sigma_{i_k} \neq \emptyset $,
which lies in $\alex^-_{i_1 \ldots i_k} \equiv \alex^-(p_{i_1})
\bigcap \ldots \bigcap \alex^-(p_{i_k}) $. Let $a \in \sigma_{i_1
\ldots i_k}$ and $\cnn \ni a$ such that $\cnn \subset \alex^-_{i_1
\ldots i_k}$. Then, $\alex^+(\Sigma) \cap N \subset \alex_{i_1 \ldots
i_k }\equiv \alex^-_{i_1 \ldots i_k} \bigcap \alex^+(\Sigma)$, implies
that $\alex_{i_1 \ldots i_k } \neq \emptyset$.  Thus every $k$-simplex
in $\cN(\mS)$ maps to a $k$-simplex in $\cN(\bI)$,
i.e. $\Psi^\ast:\cN(\bI) \rightarrow \cN(\mS)$ is onto. Since
$\alex_{i_1 \ldots i_k} \subset I^+(\Sigma)$, for every $y \in
\alex_{i_1 \ldots i_k} \ne \emptyset $, there exists an $a \in \Sigma$
such $a \prec \prec y$ and hence $a \in \alex^-_{i_1 \ldots i_k} \cap
\Sigma = \sigma_{i_1 \ldots i_k } \neq \emptyset$. Thus
$(\Psi^{\ast})^{-1}: \cN(\mS_\vl) \rightarrow \cN(\bI_\vl)$ is onto
and hence $\Psi^\ast $ is a bijection. From Theorem \ref{main}, we see
that
\begin{corollary} \label{pastset}
$\cN(\bI_\vl)$ is homotopic to $\Sigma$ and thence to $M$.  
\end{corollary}  

The cover $\mS_\vl$ of $\Sigma$ obtained from the causal structure can
therefore be used to obtain a discretisation of $\Sigma$, which is
homotopic to $\Sigma$. This provides a new way of obtaining a {\it
faithful} finitary topology from $\Sigma$ \cite{finitary} using the
geometry of the spacetime. It is tempting at this stage to speculate
on how geometrical information may be extracted from the finitary
topology.  Namely, starting from a purely topological discretisation
of $\Sigma$ using a simple open covering, if one simply assumes this
to be a convex cover $\mS_\vl$ for some choice of $\vl$, then at least
partial construction of a spacetime geometry may be possible. However,
we leave such investigations aside for now.

It is important to understand the role played by the convexity volume
$\tvl$ associated with every $\Sigma$.  For any $\vl> \tvl $, there
exist sets $\sigma_i \in \mS_\vl$ which do not satisfy \ref{nesting},
and \ref{principal} of Lemma \ref{cnn} and hence are not convex.
However, while Theorem \ref{dederham} gives a sufficient condition for
$\cN(\mS_\vl)$ to be homotopic to $\Sigma$, it is not a necessary
condition. In other words, it is possible that there exist $\vl> \tvl
$ such $\cN(\mS_\vl)$ is homologically equivalent to $\Sigma$, that
even though $\mS_\vl$ is not a convex cover, as long as its a simple
cover. Thus, at best, $\tvl$ is a lower bound for which Theorem
\ref{main} is valid.

For the purposes of the next section it will be useful to find a
slight generalisation of $\bI_\vl$ and $\mS_\vl$ to accommodate sets
associated with different $\vl$'s.  Let $\cM \equiv \{m_i
=\xi_{\vl_i}(s_i) \}$ be a collection of spacelike related points, and
$\bI \equiv \{ \alex_i' \equiv \alex(\Sigma, m_i\} $, $
\mS \equiv \{\sigma_{\vl_i}(s_i) \equiv \alex^-(\xi_{\vl_i}(s_i) \cap
\Sigma \}$, for $\vl_i < \tvl, \, \, \forall \, \, i$. Each of the
$\sigma_{\vl_i} (s_i)$ are therefore convex. Further, if the $\cM$ are
chosen such that $\mS$ is a cover of $\Sigma$, then Theorem \ref{main}
and Corollary \ref{pastset} imply that 
\begin{corollary} \label{varyingvl}
$\cN(\mS)$ and $\cN(\bI)$ are homotopic to $\Sigma$ and thence to $M$.  
\end{corollary}

We end this section with some qualitative comments on the above
foliation $\{ \Sigma_\vl \} \times \re$ of $M$. Consider the
sandwiched region $M_\vl$, with boundaries $\Sigma$ and $\Sigma_\vl$,
for either $\vl>0$ or $\vl <0$. The volume thickening has the effect
of ``smoothing out'' $\Sigma$, so that $\Sigma_\vl$ has a more
``uniformised'' or homogenised extrinsic curvature compared to
$\Sigma$.   Thus, this volume thickening has the effect of homogenising
the extrinsic curvature of the slices in the foliation, both to the
past and the future of $\Sigma$.  This is a 
``dissipative'' process analogous to a heat equation and is therefore
irreversible in general: the volume foliation of $M$ with
respect to some $\Sigma_{\vl_0}$, $\vl_0 > 0$ is not in general
equivalent to  that with respect to $\Sigma$, and the leaves of the
foliation  coincide only at $\Sigma_{\vl_0}$, so that a past volume
thickening of $\Sigma_{\vl_0}$ will in general not contain $\Sigma$ as a
slice. Such uniformising foliations may find an application in studies
of Lorentzian geometry, analogous to the Ricci flows in Riemannian
geometry \cite{ricciflow}.

\section{The Correspondence} 

We start with a causal set $C$ that faithfully embeds into a 
globally hyperbolic spacetime $(M,g)$   at density $V_c^{-1}$,  
\begin{equation}
\Phi: C \rightarrow (M,g).    
\end{equation} 
$V_c$ is the {\sl cut-off volume} which sets the discreteness scale
near which the manifold approximation of the causal set $C$ breaks
down. 

The probability of finding $n$ points in a given volume $V$ 
for a Poisson sprinkling at density $V_c^{-1}$ is 
\begin{equation} \label{poisson} 
P_n(V) = \frac{1}{n!}\exp^{-\frac{V}{V_c}} \times 
\biggl(\frac{V}{V_c}\biggr)^n,  
\end{equation}  
which implies that $\overline n = V/V_c$, with standard deviation,
$\sigma=\sqrt{{V/V_c}}$.  The relative width $\sigma/\overline n $ of
of this distribution decreases with increasing $\overline n$. Hence,
for $V\gg V_c$ not only is $P_n(V)$ appreciable only for $n\gg 1$, but
$n= \cO(V/V_c)$ with probability close to 1. This feature of the
Poisson distribution will help guide us through the main results of
this section.

In Section \ref{nerve} we constructed the nerve $\cN(\cP)$ associated
with a thickened antichain $\cT_{\hv}(\can)$. Our objective is now to
find an appropriate continuum analog $\bI$ of $\cP$ such that
$\cN(\bI)$ is homologous to $\cN(\cP)$. If $\bI$ further satisfies the
conditions of Corollary \ref{varyingvl} this would mean that
$\cN(\cP)$ is homologous to $M$.

For any \ia{} $\can$, there exists an infinity of Cauchy surfaces
containing it. This makes the discrete-continuum correlation highly
non-unique. Nevertheless, one can make a  choice
of $\can$ such that this non-uniqueness becomes unimportant to
recovering the continuum homology.

Let $\Sigma_1,\Sigma_2$ be a pair of Cauchy surfaces such that $\can
\subset \Sigma_i$, $i=1,2$. A {\sl spacing} between the $\Sigma_i$ can
be defined via a spacetime interval of the type $\alex(p_1,p_2)$,
or $\alex(q_2,q_1)$, for any $p_i,q_i \in \Sigma_i$, $i=1,2$ such that
$p_1 \prec \prec p_2$, and $q_2 \prec \prec q_1$. 
\begin{claim}
For any pair of Cauchy hypersurfaces $\Sigma_1, \Sigma_2 \supset
\can$, the spacings between them lie in $S(\can)$.
\end{claim} 
\bproof Without loss of generality, let there exist $p_i\in \Sigma_i$,
$i=1,2$ such that $p_1 \prec \prec p_2$. Then, by transitivity, for
any $q \in \alex(p_1,p_2)$, $q \nin \alex^\pm(\can)$, since the
$\Sigma_i$ are achronal. Further, if $q \in \pd \jj^\pm (\can)$, by global
hyperbolicity, this means that there exists an $a \in \can$ such that
$a \rightarrow q$ or $q \rightarrow a$. Assuming the former wlog, 
since $q \prec \prec p_2$, this means that $a \prec \prec p_2$, which
is a contradiction. Thus, $\alex(p_1,p_2) \subset S(\can)$. 
 \eproof

Since $\can$ is an \ia, $S(\can)$ must be made up entirely of {\sl
voids}, i.e., $\Phi(C) \cap S(\can) = \emptyset$. Thus, for any pair
of $\Sigma_i$'s, the spacings between them must be empty of causal set
elements. Voids of volume $V$ occur with probability
$P_{0}(V)=\exp^{-V/V_c}$. Thus, the probability for a region of volume
$V \gg V_c$ to be a spacing (and hence a void) is $\ll 1$. Given some
$\Sigma_i \supset \can$, the probability for any interval $I(p_i,q)$
to its future or $I(r,p_i)$ to its past of volume $V$ to be a spacing
is then roughly limited by the number $N$ of such independent
intervals. The probability that at least one of these intervals be a
spacing is then $NP_0(V)$, which for $N$ sufficiently small, or
$\Sigma$ sufficiently compact, is still $\ll 1$ for $V \gg V_c$

Even for a spacetime with a ``sufficiently'' compact $\Sigma$, it may
be possible, although not probable, to pick from $\Phi(C)$ an
inextendible antichain $\can$ such that the volume of the spacings 
between the $\Sigma_i \supset \can$ are not of $\cO(V_c)$. We will
henceforth only restrict to those $\can$, for which this is not the
case, i.e. the spacings are all of order $V_c$. 
This ensures that the Cauchy hypersurfaces containing such an $\can$
differ from each other by a relative volume of $\cO(V_c)$, and hence
may be considered ``equivalent'' to each other at scales $\gg V_c$.  We
may therefore pick a $\Sigma$ from this set with the largest convexity
radius $\tvl$, and directly assign $\tvl$ to $\can$.

We now explore the obvious analogy between the thickened antichain
$\cT_{\hv}(\can)$ for a likely $\can$ and the thickened region,
$M_\vl$ sandwiched between $\Sigma \supset \can$ and $\Sigma_\vl$,
where $\vl= \hv V_c$.  Corollary \ref{pastset} tells us that for $0<
\vl < \tvl$, $\cN(\bI_\vl)$ is homologous to the spacetime. While the
upper bound $\tvl$ obtains from continuum considerations, in the
discrete context, one also expects a lower bound $\vl_0$ arising from
the discreteness scale. Indeed, such a lower bound makes its
appearance in our numerical investigations.  For example, for $\vl
\sim V_c$, then there is a high probability for voids which ``cut
through'' $M_\vl$ in a swiss-cheese-like fashion. If $\cM= \{ m_i \}$
are the maximal elements of $\cT_{\hv}(\can)$, the set of shadows $\{
\sigma(m_i) \}$ on $\Sigma$ will most likely {\it not} cover $\Sigma$
for $\hv \sim 1 $ ($\vl \sim V_c$). In other words, $M_\vl$ is ``too
thin'' to be a good approximation to $\cT_\hv(\can)$ for $\hv \sim
1$. Thus $\bI \equiv \{ \alex(\Sigma, m_i ) \} $ does not satisfy the
conditions of Corollary \ref{varyingvl}, so that $\cN(\bI)$ need not
be homologous to the spacetime.  Thus, it seems necessary to impose
the condition that $\vl\gg V_c$. Clearly, this is not sufficient for
obtaining an $\cN(\bI)$ homologous to the spacetime since $\tvl$
itself may be $\cO(V_c)$ so that there exists no $\vl\gg V_c$ with
$\vl < \tvl$, and hence the conditions of Corollary \ref{varyingvl}
are not satisfied.  Hence we must further restrict to \ias{} $\can$
for which $\tvl \gg V_c$.

$M_\vl$ can also fail to be a good continuum approximation of
$\cT_\hv(\can)$, even though $\tvl > \vl \gg V_c$, since it could be
too thin in local patches.  For example, if the extrinsic curvature at
a point on $\Sigma$ is relatively large in comparison with other
points in its neighbourhood, the spacings between points on $\Sigma$
and $\Sigma_\vl$ may still be of order $V_c$ even for $\vl\gg
V_c$. Fig \ref{spacing.fig} provides an illustration of this. Let us
define
\begin{equation} 
\alpha_x(\vl) \equiv  \sup_{y \in \Sigma_\vl} \vol(\alex(x,y)), \qquad x \in
\Sigma, 
\end{equation} 
which is the size of the largest spacetime interval between a given $x
\in \Sigma$ and any $y \in \Sigma_\vl$.  If $\alpha(x) \sim V_c$, then
$M_\vl$ is once again too thin at $x$ and not a good continuum
approximation to $\cT_{\hv}(\can)$. Again, the probability of a void
at $x$ which cuts through $M_\vl$, is high.  On the other hand, if
$\alpha_x(\vl) \gg V_c$ (which implies $\vl \gg V_c$) for all $x \in
\Sigma$, then such voids are less probable.  We may therefore define
the lower bound $\vl_0$ on $\vl$ such that $\forall \, \, \vl> \vl_0$,
$\alpha_x(\vl) \gg V_c$, $\, \forall \, x \in \Sigma$
\begin{figure}[ht]
\centering
\resizebox{3.5in}{!}{\includegraphics{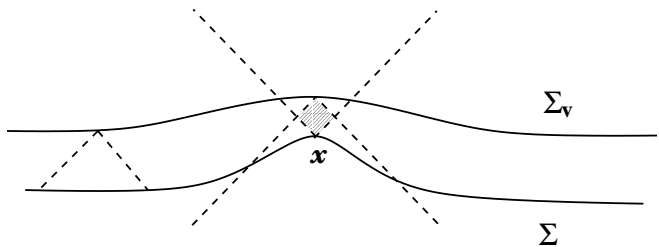}}
\vspace{0.5cm}
\caption{{\small The largest interval  between a given  $x \in
    \Sigma$ and a  point in $\Sigma_\vl$.}}\label{spacing.fig}
\end{figure}  

In order to obtain the high probability discrete-continuum
correspondence for the nerves, we will find it necessary to impose a
very strong separation of scales, $ V_c \ll \vl_0 \ll \vl \ll
\tvl$. This may be too stringent a requirement in practice, but it
allows us to make our arguments in as simple and general a language as
possible.  It is useful to see if such a separation of scales lies
within physically reasonable bounds. We can assume that the continuum
description is valid at least down to length scales $l_c= 10^{-25 } \mm$,
which gives us a cut-off volume of $V_c=10^{-100} \mm^4$. Let
$ \alpha_x(\vl_0)\sim 10^{12} V_c$ which has an associated length
scale of $ 10^{-22}\mm$. A conservative choice for $\tvl$, which
corresponds to the shortest scales at which spatial intrinsic and
extrinsic curvature effects become important, is the nuclear scale
$\sim 10^{40} V_c$. Thus, a separation of scales $ 10^{40}V_c\gg \vl
\gg 10^{12}V_c \gg V_c $ is possible within reasonable physical
bounds, i.e., far above the Planck scale and far below the nuclear
scale.

We have therefore placed fairly stringent requirements on the \ia{}
$\can$ with respect to its faithful embedding into the spacetime
$(M,g)$ at a {\it fixed} density $V_c^{-1}$. The obvious question is
how to identify such $\can$ in $C$ without reference to the embedding
$\Phi(C)$ in $(M,g)$. However, this requires a better understanding of
intrinsic geometrical information in the causal set than we presently
have and is under current investigation \cite{us}.  We instead adopt
the following, mathematically equivalent, approach. Starting with a
given $(M,g)$, we allow the sprinkling density to vary such that any
$M_\vl$ with $\vl < \tvl $ satisfies the condition $\vl > \vl_0 \gg
V_c$.  Thus, we pick only those $\can$'s from $\Phi(C)$ satisfying this
separation of scales.

Given an $\can$ satisfying these above conditions, with associated
$\tvl = \thv V_c , \vl_0= \hv_0 V_c $, consider its thickening,
$\cT_{\hv}(\can)$, where  $ \thv \gg \hv \gg \hv_0 \gg
1$.   For any $y \in \jj^+(\Sigma)$, define $\vl(y)
\equiv \vol(\alex(\Sigma,y))$ and $\nsf(y) \equiv |\ifut(\can) \cap
\ipast(y)|$.  

We now list  probabilities for certain occurrences which will
subsequently be used to quantify the ``high probability''
correspondence between the continuum and the discrete homologies.  The
probability for a void of volume $V' \gg V_c$ is
\begin{equation} \label{pzero}
\pP_0 \equiv \exp^{-V'/V_c} \ll 1
\end{equation} 
The probability for a region of volume $\vl_0$ to have $\hv $ 
sprinkled points with $ \hv \gg
\hv_0\gg 1 $ is
\begin{equation}\label{pone}
\pP_1 \equiv \sum_{k=\hv}^\infty \frac{1}{k!}\exp^{-\hv_0}
(\hv_0)^k = 1 - \frac{\Gamma(\hv, \hv_0)}{\Gamma(\hv)}
\ll 1,  
\end{equation}  
where $\Gamma(x,y)$ is the incomplete Gamma function.  This is the
probability of an ``overstuffed'' region.  The probability for a
region of volume $\tvl = \thv V_c$ to contain $\hv$ sprinkled points,
with $1\ll \hv \ll \thv$ is
\begin{equation}\label{ptwo} 
\pP_2 \equiv \sum_{k=0}^\hv \frac{1}{k!}\exp^{-\thv_0}
(\thv_0)^k = 
\frac{\Gamma(\hv+1, \thv)}{\Gamma(\hv +1)} \ll 1.  
\end{equation}   
This is the probability of an ``understuffed'' region.  One will need
to calculate the total probability for at least one such improbable
situation (say a void of volume $ \gg V_c$) to occur {\it anywhere} in
the thickened spacetime region $M_{\vl_0}$. Since the voids can
overlap, one can put an upper bound on the probability by considering
non-overlapping regions, which can be constructed from the
non-overlapping shadows $\sigma_i$ from events on $\Sigma_{\vl_0}$.
For a given $\can$ and associated $\Sigma$, the volume of a spacetime
interval $\alex(\Sigma,p)$ for any $p \in \Sigma_{\vl_0}$ is
$\vl_0$. Since $\vol(M_{\vl_0})$ is roughly $ \sim \vl_0^{1/\dimm}
\vol(\Sigma)$, where $\dimm$ is the spacetime dimension
\footnote{Since these intervals lie in CNN's where the deviation from
flatness is small, this suffices to give a rough order of magnitude
estimate, which is all we require.}, the number $\nN$ of such
non-overlapping regions in $M_{\vl_0}$ is
\begin{equation} 
\nN \sim \vol(\Sigma) \vl_0^{-\frac{(\dimm-1)}{\dimm}}.
\end{equation}
For $\vl > \vl_0$, the corresponding number of independent regions,
$\nN_\vl < \nN$ and hence we may use $\nN$ as an upper bound for all
such $\vl$. Instead of encumbering ourselves with notation, we will
then simply use $\nN$ itself as an adequate measure of the number of
non-overlapping regions, even for $\vl \gg \vl_0$.  
If $\nN$ is very large, then $\Sigma$ is correspondingly large and the
total probability for a rare event may be non-negligible. Thus, we
quantify our requirement for $\Sigma$ to be  {\sl sufficiently
  compact} by 
\begin{equation} \label{compact}
\nN \pP_0 \ll 1, \qquad  \nN \pP_1 \ll 1, \qquad  \nN \pP_2 \ll 1. 
\end{equation}  
It is instructive at this point to check how restrictive this is for
our present universe.  If we take $\can$ to lie in a homogeneous
isotropic $\Sigma$, assuming $V_c =V_p$, roughly, $10^{180} \leq \nN \leq
10^{240}$.  The probability that there are no voids of volume $\gg
10^4 V_p$, say, is then $< 10^{240} \times \exp^{-10000} \sim
\exp^{-9447} \ll 1$.

We now demonstrate some results  which will simplify the proof
of our final  Lemma. 
\begin{claim} \label{nogap} 
For any $x \in \Sigma$, there exists at least one point in
$\alex^+(x)$ which belongs to $\cT_{\hv}(\can)$ with high
probability. 
\end{claim} 
\bproof For any $x \in \Sigma$, $\alpha_x(\vl_0) \gg V_c$, so that 
$\vol (\alex(x, \Sigma_{\vl_0}))\gg V_c$.  The probability of
finding at least one sprinkled point in $\alex(x, \Sigma_{\vl_0})$ is
therefore $1-\pP_0$, where $\pP_0 \ll 1$. The probability of finding at
least one void in $M_{\vl_0}$ of this kind is then $\sim \nN \pP_0 \ll
1$.    

For any $p \in \alex(x, \Sigma_{\vl_0})$,  $\vl(p) < \vl_0$. The
probability that it has $\hv(p) > \hv \gg \hv_0 $ is less than $\pP_1
\ll 1$. Hence, the probability for at least one of the $\nN$
non-overlapping intervals to have such a point is less than $\nN
\pP_1$. The total (joint) probability for this occurrence is then 
$(1-\nN \pP_0)\times \nN \pP_1 \sim \nN \pP_1 \ll 1$. Conversely, the probability
for every $x \in \Sigma$ to contain at least one point in its future
belonging to $\cT_{\hv}(\can)$ is $1 - \nN \pP_1 \sim 1$. 
\eproof 

\noindent {\bf Note:} The above claim also holds for any $y \in
\alex^+(\Sigma)$ such that $\vl(y) = \cO(V_c)$, since $\alpha_y(\vl_0)
 \gg V_c$.

\begin{claim} \label{maximal} 
Let $\cM$ be the set of maximal elements in $\cT_{\hv}(\can)$. Then
with high probability, none of the elements $e_i \in \Phi(C)$ with
$\vl(e_i) = \cO(V_c)$ belong to $\cM$.  
\end{claim}

\bproof Let $e_1 \in Fut(\can)$ such that $\vl(e_1)=\cO(V_c)$. Then,
$\vl(e_1) \ll \vl_0$ and hence $\alex(e_1, \Sigma_{\vl_0}) \gg
V_c$. Then $\alex(e_1, \Sigma_{\vl_0}) \cap \cT_\hv(\can) = \emptyset$
occurs with probability $\sim \pP_1 \ll 1$ from Claim
\ref{nogap}. Therefore $e_1$ is not a maximal element with probability
$> 1 -\pP_1$. The number of $e_i$ with $\vl(e_i)=\cO(V_c)$ $
\Rightarrow \alex(e_i, \Sigma_{\vl_0}) \gg V_c$ which are
non-overlapping is $\sim \nN$, and hence the joint probability that
none of them is a maximal element is $\sim 1-\nN \pP_1 \sim 1$.
\eproof

\begin{claim} \label{shadowcover} 
Every $x \in \Sigma$ lies in a shadow $\sigma_i \equiv
\alex^-(m_i)\cap \Sigma$ with high probability. 
\end{claim} 
\bproof The proof of this is immediate from Claim \ref{nogap}. Namely,
the probability that there exists at least one $x \in \Sigma $ which
does not lie in a shadow is $\nN \pP_1$ and hence with probability
$1-\nN \pP_1 \sim 1 $ every $x$ lies in a shadow. \eproof.  

\begin{claim} \label{homotopy}
Let $\bI= \{ \alex_i \}$, with $\alex_i \equiv \alex(\Sigma, m_i)$. 
$\cN(\bI)$ is homotopic to $M$ with high probability.  
\end{claim} 
\bproof From Claim \ref{shadowcover} with probability $1-\nN \pP_1$, $\mS
\equiv \{ \sigma_i \}$ is a cover of $\Sigma$. Moreover, with
probability $1-\nN \pP_2$, $\vl(m_i) < \tvl$, so that the joint
probability for $\mS$ to be a convex cover of $\Sigma$ is $ \sim 1-\nN
(\pP_1+ \pP_2)$. Thus, from  Corollary \ref{varyingvl},  $\cN(\mS)$
and $\cN(\bI)$ are both homotopic to $M$ with probability $ \sim 1-\nN
(\pP_1+ \pP_2) \sim 1$. \eproof.

We are now in a position to begin the discrete-continuum comparison.

\begin{claim}\label{vertices}  
Define the discrete collection of sets $\cP= \{ P_i\}$, $P_i \equiv
\fut(\can) \cap \ipast(m_i)$\footnote{ Note that the $P_i$ do not
include the elements of $A_i$ in this definition, and hence differ
from the $P_i$ of Section \ref{nerve}. The reason to adopt this
modified choice comes from the fact that unlike in the continuum,
there is no natural distinction between elements which are causally
related and those that are strictly chronologically related in a
causal set.  In particular, Claim \ref{subcomp} would not be valid
without this modification.}. There is a one-one and onto map from
$\bI$ to $\cP$. Hence the vertices of $\cN(\cP)$ and $\cN(\bI)$ are in
one-one correspondence.
\end{claim}  
\bproof 
For every $m_i$ there exists a unique $P_i$, 
since the set of maximal points $\cM$ form an antichain (which need
not be inextendible) and hence $m_i \in P_j$ iff $m_i=m_j$. Moreover,
because the spacetime is distinguishing, every $m_i$ has a unique
$\alex^-(m_i)$ and hence a unique $\alex_i$. \eproof.

This bijection between the vertices of $\cN(\cP)$ and $\cN(\bI)$
allows us to label the vertices as $m_0, m_1, \ldots
m_\Dimm$. Therefore the largest possible dimension of a simplex in
$\cN(\bI) $ or $\cN(\cP)$ is $\Dimm$.

\begin{claim}\label{subcomp} 
$\cN(\cP)$ is a subcomplex of $\cN(\bI)$.  
\end{claim} 
\bproof For any $p \in P_{i_1\ldots i_k} \equiv P_{i_1} \cap \ldots
\cap P_{i_k} \neq \emptyset $, $p \prec m_{i_j} $ for all $j = 1,
\ldots k$. Since $p \in \alex^+(\Sigma)$, $\, \exists \, x \in
\alex^+(\Sigma)$ such that $p \in \alex^+(x)$\footnote{Without the
  modified definition of the $P_i$ this would no longer be true. For
  example, if the intersection is non-empty only {\it on} the light cones
  and on $\can$, then any $p \in \can$ and does not lie in an
  intersection of $\alex_i$s.}, so that $ m_{i_j} \in
\alex^+(x) $, for all $j = 1, \ldots k$, and hence $p \in \alex_{i_1
\ldots i_k} \equiv \alex_{i_1} \cap \ldots \cap \alex_{i_k} \neq \emptyset $.
\eproof

However, $\cN(\bI)$ is {\it not} in general a subcomplex of $\cN(\cP)$
since there may exist non-empty intersections $\alex_{i_1 \ldots i_k} \neq
\emptyset$ with $\vol(\bigcap_{i \in I} \bI_i) = \cO(V_c)$. There is then a
high probability for such intersections to be voids, i.e., $P_{i_1
\ldots i_k}=\emptyset$. If this is the case, then there is no
$k$-simplex in $\cN(\cP)$ which maps to this particular $k$-simplex in
$\cN(\bI)$. In other words, $\cN(\cP)$ and $\cN(\bI)$ need not be  homotopic
to each other. 

Despite this, we now show that $\cN(\cP)$ and $\cN(\bI)$ are
homologous to each other with high probability, since $\cN(\cP)$ is an
{\sl adequate subcomplex} of $\cN(\bI)$ with high probability.  We
begin by establishing some notation and reminding the reader of
some basics of algebraic topology \cite{rotman}.

We begin by putting an orientation on the simplicial complex
$\cN(\bI)$, so that the set of vertices $\{ m_i\} = \cM $ are ordered
as $m_0 \rightarrow m_1 \rightarrow \ldots \rightarrow m_\Dimm$.  Let
$C_q(\NI)$ be the free abelian group with basis $ \cB_q \equiv
\{\hbq^{(\al)}\}$, where the $\hbq^{(\al)}=(m_{s_0}, \ldots, m_{s_q}
)$ are $q$-simplices in $\NI$, such that $(m_{s_{\pi(0)}}, \ldots
m_{s_{\pi(q)}}) = \mathrm{sign}(\pi) \, \, (m_{s_0}, \ldots m_{s_q}) $
for $\pi$ a permutation of the set $\{ 0, 1 , \ldots , q\}$.  Elements
of $C_q(\NI)$ are referred to as $q$-chains.  For $q> \Dimm$,
$C_q(\NI) = \emptyset$. $\bZ_q(\NI) \subset C_q(\NI)$ are the
simplicial $q$-cycles and $\bB_q(\NI) \subset C_q(\NI)$ the simplicial
$q$-boundaries, so that the $q$th homology group $\Hom_q(\NI) \equiv
\bZ_q(\NI)/\bB_q(\NI)$.  It will be convenient to use the shortened
notation $K \equiv \NI$, $C_q \equiv C_q(\NI)$, $\bZ_q \equiv
\bZ_q(\NI)$, $\bB_q \equiv \bB_q(\NI)$ and denote with primes the
associated sets for the simplicial complex $K' \equiv \NP$.

Now, a $q$-simplex $(m_{s_0}, \ldots, m_{s_q} )$ is an element of
$\cB_q$ iff the intersections $\alex_{s_0\ldots s_q}\neq \emptyset$, and
similarly, $(m_{s_0'}, \ldots, m_{s_q'} )$ is an element of $\cB_q'$
iff $P_{\al_0'\ldots  \al_q'}\neq \emptyset$. We will use the notation
$(s_0, \ldots,  s_q) \equiv (m_{s_0}, \ldots, m_{s_q} )$ and $ [s_0,
  \ldots, s_q] \equiv \alex_{s_0\ldots s_q}$, which also helps us
switch more easily from sets to simplices and back. As in standard
notation a $q-1$-simplex $(s_0, \ldots, \widehat{s}_i,  \ldots,   s_q)
$ can be obtained from a $q$-simplex $(s_0, \ldots,   s_q)$ by simply
dropping the $i$th vertex. 

A subcomplex $K'$ of a simplicial complex $K$ is said to be {\sl
adequate} if for all $q\geq 0$,
\begin{enumerate}  
\item \label{one} if $z \in \bZ_q$, then there exists a $z' \in
\bZ_q'$ with $z-z' \in \bB_q$ and 
\item  \label{two} if $z' \in \bZ_q'$ and $z' = \pd c$
for some $c \in C_{q+1}$, then there exists a $c' \in C_{q+1}'$ with
$z'=\pd c'$.
\end{enumerate} 
\begin{lemma}{\bf (Rotman)}  If $K'$ is an adequate subcomplex of
$K$, then for every $q$, the map $z' + \bB_q' \mapsto z' + \bB_q$ is
an isomorphism 
\begin{equation}
\Hom_q(K')  \simeq \Hom_q(K). 
\end{equation} \label{rotman} 
\end{lemma}
Thus, our task is reduced to showing that $\NP$ is an adequate
subcomplex of $\NI$.  

We will begin by defining a {\sl growth} of a $q+1$-chain $\tau$ from
a given $q$-simplex, $\cF_0 \equiv (s_0, \ldots, s_q)$.  We describe
this construction in some detail, since it is crucial to the proof of
the main results of this section. Starting with some $\cF_0$, let
there exist an $s_{q+1}$ such that there exists a non-empty
$q+1$-simplex 
\begin{equation} 
\tau_1 \equiv (s_0, \ldots , s_q, s_{q+1}) \neq \emptyset. 
\end{equation} 
We have thus ``grown'' a $q+1$-chain (in this case simply a
$q+1$-simplex) from $\cF_0$. The boundary of
$\tau_1$ then consists of the set of $q$-simplices, 
\begin{equation} \label{Fone}
\cF_1 \equiv \{(s_0,
\ldots , \widehat{s}_i, \ldots , s_{q+1}) \},  \forall i \in [0,q],  
\end{equation}
and the original $q$-simplex $\cF_0$. We may stop the growth here.

However, if for some $i_1\in [0,q]$, let there exist an $s_{q+2}^{(i_1)} \neq
s_{i_1}$ such that
\begin{equation} 
\tau_2^{(i_1)}
\equiv (s_0, \ldots , \widehat{s}_{i_1}, \ldots , s_{q+1},s_{q+2}^{(i_1)}
) \neq \emptyset, 
\end{equation}   
then the growth can be continued, to obtain a larger $q+1$-chain
containing $\tau_1$.  The $q+1$-simplex $\tau_2^{(i_1)}$ consists of the set of
$q$-simplices on its boundary
\begin{equation}\label{Ftwo}  
\cF_2^{(i_1)} \equiv \{ (s_0, \ldots , \widehat{s}_{i_1}, \ldots,
\widehat{s}_{j} , \ldots, s_{q+1},s_{q+2}^{(i_1)}) \}, \quad \forall  j \in [0,
\ldots \widehat{i}_1, \ldots q+1],
\end{equation} 
along with $(s_0, \ldots, \widehat{s}_{i_1}, \ldots ,
s_{q+1},\widehat{s}_{q+2}^{(i_1)})$.  $\tau_1$ and $\tau_2^{(i_1)}$
then share the  single $q$-simplex along their boundary,
\begin{equation} 
\ssigma^{(i_1)}\equiv (s_0, \ldots, \widehat{s}_{i_1}, \ldots,
s_{q+1},\widehat{s}_{q+2}^{(i_1)} ),  
\end{equation} 
and the boundary of the $q+1$-chain $\tau_1+\tau_2^{(i_1)}$ 
consists of the set of $q$-simplices 
\begin{equation}  \label{Bone} 
\cF_0\bigcup \cF_1 \bigcup \cF_2^{(i_1)} - \ssigma^{(i_1)}.
\end{equation} 
Again, we could stop the growth here, along the $i_1$th branch.

If for a given $i_1$, there is an   $i_2 \in [ 0, \ldots,
  \widehat{i}_1, \ldots q+1] $ such that there exists an 
  $s_{q+3}^{(i_1i_2)} \neq s_{i_1}, s_{i_2}$ such that   
\begin{equation} 
\tau_3^{(i_1i_2)}
\equiv (s_0, \ldots \widehat{s}_{i_1},  \ldots ,\widehat{s}_{i_2},
\ldots   s_{q+1},s_{q+2}^{(i_1)}, s_{q+3}^{(i_1i_2)}) \neq \emptyset,  
\end{equation} 
then we may continue the growth in the $i_1i_2$ branch. The
$q+1$-simplex $\tau_3^{(i_1i_2)}$ consists of the set of $q$-simplices
on its boundary
\begin{equation} 
\cF_3^{(i_1i_2)} \equiv \{ (s_0, \ldots,  \widehat{s}_{i_1},  \ldots,
\widehat{s}_{i_2} , \ldots, \widehat{s}_k, \ldots,  s_{q+3}^{(i_1i_2)})
\}, \quad \forall  j \in [0, \ldots,  \widehat{i}_1, \ldots,
  \widehat{i}_2,  \ldots,  q+2],
\end{equation} 
along with $(s_0, \ldots, \widehat{s}_{i_1}, \ldots, \widehat{s}_{i_2},
\ldots, \widehat{s}_{q+3}^{(i_1i_2)})$. 
$\tau_2^{(i_1)}$ and $\tau_3^{(i_1i_2)}$ then share a single $q$-simplex along
their boundary, 
\begin{equation} 
\ssigma^{(i_1i_2)}\equiv (s_0, \ldots \widehat{s}_{i_1} \ldots,
\widehat{s}_{i_2} \ldots \widehat{s}_{q+3}^{(i_1i_2)}).  
\end{equation} 
The boundary of the $q+1$-chain $\tau_1+\tau_2^{(i_1)} +
\tau_3^{(i_1i_2i_3)}$ then contains the set of $q$-simplices
\begin{equation}  
\cF_0\bigcup \cF_1 \bigcup \cF_2^{(i_1)} \bigcup \cF_3^{(i_1i_2)} - \ssigma .
\end{equation} 

In this manner, we may carry on the growth of the $q+1$ chain along
each branch upto the point desired, or until it cannot be grown
further. For a finite simplicial complex this will end in a finite
number of steps. The final $q+1$-chain is 
\begin{equation}\label{tau} 
\tau \equiv \tau_1 + \biggl[\;\sum_{i_1 \in G_1} \biggl[\tau_2^{(i_1)} + \sum_{i_2
    \in G_{12}} \biggl[ \tau_3^{(i_1i_2)} \ldots \biggr]  \biggr] \biggr]  
\end{equation} 
whose boundary contains the set of $q$-simplices 
\begin{equation}\label{cF} 
\cF \equiv \cF_0 \bigcup \biggl[ \cF_1 \biggl[\bigcup_{i_1 \in G_1}
  \cF_2^{(i_1)}\biggl[\bigcup_{i_2 \in G_{12}} \cF_3^{(i_1i_2)}  \ldots
  \biggr]\biggr]\biggr] - \ssigma, 
\end{equation} 
where 
\begin{equation} 
\ssigma \equiv \biggl[\bigcup_{i_1 \in G_1}
  \ssigma^{(i_1)} \biggl[\bigcup_{i_2 \in G_{12}}
  \ssigma^{(i_1i_2)} \ldots  \biggr] \biggr], 
\end{equation} 
and $G_1$ is the set of $i_1$, $G_{12}$ the set
of $i_2$'s in the $i_1$ branch, etc. Important in this growth, 
is the set $\ssigma$ which has been ``removed'' from the boundary of
$\tau$. We illustrate for the simple case of the growth of a
$2$-chain from a $1$-simplex in Figure \ref{growth.fig}. 
\begin{figure}[ht]
\centering
\resizebox{3.5in}{!}{\includegraphics{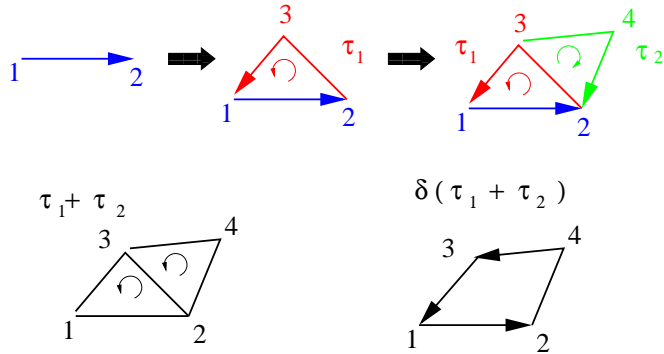}}
\vspace{0.5cm}
\caption{{\small The growth of a $2$-chain from a
$1$-simplex.}}\label{growth.fig}
\end{figure}  
We will now use this  growth process to eliminate the set $\ssigma$ of
$q$-simplices which have volume $\cO(V_c)$, or are ``thin''.  
We will say that $[s_0, \ldots, s_q] \neq \emptyset$ is ``thin'' if
$\forall x \in [s_0, \ldots, s_q]$, $\vl(x) = \cO(V_c)$ and ``fat'' if
there exists a maximal element $x \in \overline{[s_0, \ldots, s_q]}$
such that $\vl(x) \gg V_c$.

\begin{claim}\label{fat}  
For every $\cF_0 \equiv (s_0, \ldots, s_q)$ such that $\vol[s_0,
\ldots, s_q] = \cO(V_c) \neq \emptyset$, there exists, with high
probability, a $q+1$-chain $\tau$ grown from $\cF_0$ as in (\ref{tau})
with boundary made up of the set of $q$-simplices $\cF$ as in
(\ref{cF}), all of which correspond to ``fat'' sets, save $\cF_0$
itself.
\end{claim} 
\bproof Since $[s_0, \ldots, s_q]$ is ``thin'', with probability
$P(q+1) \sim 1- \pP_1$ there exists an $m_{s_{q+1}}$ which lies in the
future of a maximal event in $\overline{[s_0, \ldots, s_q]}$. This
follows from Claim \ref{nogap}.  By set inclusion, $ [s_0, \ldots,
s_q, s_{q+1}] \subseteq [s_0, \ldots, s_q]$ and moreover, $[s_0,
\ldots, s_q, s_{q+1}] \neq \emptyset$ by transitivity, so that
\begin{equation} 
\tau_1= (s_0, \ldots, s_q, s_{q+1}) \neq \emptyset. 
\end{equation} 
This is the first stage of the growth process. Now, if there exists no
$i_1$ such that $[s_0, \ldots, \widehat{s}_{i_1}, \ldots, s_{q+1}]$ is
thin, then we may the stop the growth process. The boundary of
$\tau\equiv \tau_1$ is made up of the set of $q$-simplices $\cF_1 \cup
\cF_0$ (\ref{Fone}) which are all fat, save $\cF_0$.

Assume otherwise, i.e.,  let there be an $i_1 \neq q+1$ such that
$[s_0, \ldots, \widehat{s}_{i_1}, \ldots, s_{q+1}]$ is
thin.  With probability $P(q+2) \sim 1-\pP_1$, there exists an
$m_{s_{q+2}}^{(i_1)}$  which lies to the future of a maximal event 
of $\overline{[ s_0, \ldots, \widehat{s}_{i_1}, \ldots, s_{q+1}]}$,
such that   
\begin{equation} 
[s_0, \ldots, \widehat{s}_{i_1}, \ldots, 
s_{q+1}] \supseteq [s_0, \ldots,\widehat{s}_{i_1}, \ldots,
  s_{q+2}^{(i_1)}] \neq \emptyset,  
\end{equation}   
so that 
\begin{equation} 
\tau_2^{(i_1)}= (s_0, \ldots,\widehat{s}_{i_1}, \ldots, s_{q+2}^{(i_1)})
\neq \emptyset.  
\end{equation} 
Now, choose the maximal event 
such that $m_{s_{q+2}}^{(i_1)} \neq
m_{s_{i_1}}$. If such a choice of $m_{s_{q+2}}^{(i_1)}$ is not possible,
then we have reached the end of the growth. Thus, we prevent ourselves
from picking up a vertex that we previously dropped along any given
branch of the growth.  Then $\tau \equiv \tau_1 + \tau_2^{(i_1)}$ has
a boundary made up of the simplices (\ref{Bone}). 

The total probability for $m_{s_{q+2}}^{(i_1)} $ and $m_{s_{q+1}}$ to
both occur is then bounded from below by $P(q+1)P(q+2) \sim 1 - 2
\pP_1$. 

Again, if there exists no $i_2$ such that $[s_0, \ldots,
\widehat{s}_{i_1}, \ldots, \widehat{s}_{i_2}, \ldots, s_{q+2}^{(i_1)}]$
is thin, then we may again stop the growth process and $\tau$ is the
$q+1$-chain $\tau_1+ \tau_2^{(i_1)}$ whose boundary is made up of the
set of $q$-simplices (\ref{Bone}),  
all of which are fat, save $\cF_0$. 

Since there are only a finite number of elements of $\cM$, this
process must stop in a finite number of steps.  If the end of the
growth along any of the branches yields a final $m_{s_{q+k}}^{(i_1i_2
\ldots i_r)}$ such that there exist sets $[s_0, \ldots,
\widehat{s}_{i_1}, \ldots \widehat{s}_{i_r}, , s_{q+r}^{(i_1 \ldots
i_r)}]$ which are {\it not} fat, then this occurs with probability
$\pP_1 \ll 1$.  Thus, with probability greater than $1- r \pP_1 \sim 1$, a
given ``final'' intersection along a branch $i_1i_2\ldots i_r$, is
fat. Since there are a maximum of $\nN$ non-overlapping regions, this
is true along all of the branches with probability $ > 1- \nN \pP_1 \sim
1$. 

The end result of this growth thus yields a $q+1$ chain $\tau$
(\ref{tau}) whose boundary is a collection of $q$-simplices
(\ref{cF}), which are all fat save $\cF_0$, with high probability.  
\eproof

\begin{lemma} 
There exists an isomorphism $\Hom_q(\NP) \simeq \Hom_q(M)$ with high
probability. 
\end{lemma} 

\bproof Let $K' \equiv \NP$ and $K \equiv \NI$.  $C_0$ is isomorphic to
$C_0'$ since the vertices of the complexes $K$ and $K'$ are the same,
by Claim \ref{vertices}.  Thus, $\cB_0 \equiv \cB_0'$ and any $z \in
\bZ_0$ lies in $\bZ_0'$ and vice versa, and every $z \in \bB_0$ lies
in $\bB_0'$, thus satisfying the two requirements for an adequate
subcomplex, when $q=0$. Henceforth we assume that $q>0$. We show one
by one that both requirements are satisfied by $\NP$ with high
probability.  \\

(i) If $z_1 \in \bZ_q$ and $z_1 \in \bB_q$, then for  $z' =
\emptyset$, $z_1-z' \in B_q$ and we're done.  Also, if $z_1
\in Z_q'$, then putting $z'=z_1$ gives us $z_1 - z' = \emptyset \in \bB_q$
and again, we're done.  

Therefore, we only need consider $z_1 \in Z_q$ such that $z_1 \nin
B_q$, $ z_1 \nin Z_q'$.  $\cB_q' \subset \cB_q$ is then a strict
inclusion. Let $z_1 = \sum_{\al=1}^r \mu_\al \hbq^{(\al)}$.  Since
$z_1 \nin Z_q'$, there exists a $\mu_{\al'} \neq 0$, such that
$\hbq^{(\al')} \nin \cB_q'$.  Now, $\hbq^{(\al')} \equiv (m_{s_0},
\ldots, m_{s_q} )$ is such that $[m_{s_0}, \ldots, m_{s_q}] \neq
\emptyset$ is a void. If $[m_{s_0}, \ldots, m_{s_q}]$ is fat, then one
has a void of volume $V \gg V_c$ which occurs with probability $\pP_0
\ll 1$ and is hence unlikely. The probability that there are no voids
of volume $V\gg V_c$ to the future of $\Sigma$ is then $\sim 1- \nN
\pP_0 \sim 1$. Hence, with probability $> 1- \nN \pP_0$ any such
$[m_{s_0}, \ldots, m_{s_q}]$ is thin. From Claim \ref{fat}, then with
probability greater than $1-\nN \pP_1$, there exists a $q+1$-chain
$\tau$ whose boundary $\pd \tau$ is made up of fat $q$-simplices, save
for $\hbq^{(\al')} $. Moreover, any fat $q$-simplex in $\cB_q$ is,
with probability $1- \pP_0$, a basis element of $\cB_q'$. Since these
sets are not independent, the probability for all the fat
$q$-simplices to be basis elements of $\cB_q'$ is bounded from below
by $1- r' \pP_0$, where $r'$ is the number of all $\mu_\al \neq 0$
save one, and is such that $r' \leq \nN$. Thus, for any basis element
$\hbq^{(\al')} \nin \cB_q'$, with probability $1-\nN (\pP_0 + \pP_1) -
r' \pP_0 \sim 1$ we can associate a $q+1$-chain $\tau$ whose boundary
$\pd \tau$ is made up of basis elements that belong to $\cB_q'$ save
for $\hbq^{(\al')} $.

Then $z_2 =z_1 - (-1)^{q+1} \mu_{\al'} \pd \tau \in \bZ_q$ and $z_1 -
z_2 \in \bB_q$. Moreover, writing $z_2 = \sum_{\al=1}^r \nu_\al
\hbq^{(\al)}$, gives $\nu_{\al'} =0$.  If this was the only basis
element in $z_1$ not in $\cB_q'$, then we're done. We may thus ``weed
out'' all the thin intersections in $z_1$ iteratively, until we
finally get a $z_k = \sum_{\al=1}^r \mu_\al' \hbq^{(\al)}$ such that
all for all $\mu_{\al} \neq 0$, the $\hbq^{(\al)}$ are fat.

Then, with probability of at least $1 - \nN (\pP_1 + 2\pP_0) \sim
1$, for every $z_1 \in \bZ_q$ there is a $z' \in \bZ_q'$ such that
$z_1 - z' \in \bB_q$.

(ii) Let $z \in \bZ_q'$ and $z \in \bB_q$, i.e. $z =\pd c_1$ for some
$c_1 \in C_{q+1}$. If $c_1 \in C_{q+1}'$, then we're done. Let us
assume otherwise, i.e. $c_1 \nin C_{q+1'}'$. Then $\cB_{q+1}' \subset
\cB_{q+1}$ is a strict inclusion.

Let $c_1= \sum_{\al=1}^{r} \mu_\al \hb_{q+1}^{(\al)} \in
C_{q+1}$. Then there exists a $\mu_{\al'} \neq 0$ such that
$\hb_{q+1}^{(\al')} \nin \cB_q' $.  For a given $\hb_{q+1}^{(\al')}
\equiv (m_{s_0}, \ldots, m_{s_{q+1}} )$, according to Claim \ref{fat},
with probability $1- \nN \pP_1$, there exists a $q+2$-chain $\tau$ whose
boundary $\pd \tau$ is made up of fat $q+1$-simplices, save for
$\hb_{q+1}^{(\al')}$.

Let $c_2 = c_1 - (-1)^{q+2}\mu_{\al'} \pd \tau \in C_{q+1}$. Clearly,
$\pd c_1 = \pd c_2$. Hence, as in (i) the thin simplices in $c_1$ may
be weeded out until one obtains a $c'$ made up of only fat basis
elements.

The analysis is similar to that in (i) and hence with probability
greater than $1 - \nN (\pP_1 + 2 \pP_0) \sim 1$, for every $z \in
\bZ_q'$ $z = \pd c_1$ such that $c_1 \in C_{q+1}$ there is a $c' \in
C_{q+1}'$ such that $z=\pd c'$.

Thus, $\cN(\cP)$ is an adequate subcomplex of $\cN(\bI)$ and therefore
by Lemma \ref{rotman},  $\Hom(\cN(\cP)) $ $ \simeq \Hom(\cN(\bI))$ 
with probability
$1 - \nN (\pP_1 + 2 \pP_0) \sim 1$. Moreover, by Claim \ref{homotopy}
$\Hom(\cN(\bI))$$ \simeq \Hom(M)$ with probability  $\sim 1-\nN
(\pP_1+ \pP_2)$. Thus, the joint probability for $\Hom(\cN(\cP))$ $  \simeq
\Hom(M)$ is $\sim 1 - \nN(2\pP_1 + \pP_2 +  2 \pP_0) \sim 1$.  
\eproof.

\section{Conclusions}

What can we conclude about the Hauptvermutung?  For an appropriate
\ia{} $\can\subset C$, our result states that there exists a wide
range of thickenings $\cT_{\hv}(\can)$, $\hv\gg 1$ for which
$H_q(\NP)$ is isomorphic to $H_q(M)$ with high probability. Thus, it
is crucial to be able to identify such an \ia{} in $C$. {\it Assuming}
that it exists, and is unique, we have proved a  weak form of the
\hpt{}, i.e., if $\Phi:C \rightarrow M$ and $\Phi':C \rightarrow M'$
are two possible faithful embeddings, then $H_q(M) \simeq H_q(M')$.

The problem therefore lies in how to identify an $\can \subset C$ with
which to perform such a test, i.e., one for which $\Phi(\can) \subset
M$ is such that the required separation of scales may be satisfied.
Roughly speaking, for causets that embed into compact spacetimes, the
largest convexity volume $\tvl$ might be expected for a Cauchy
hypersurface with the smallest intrinsic and extrinsic curvatures.  An
$\can$ that would lie in such a hypersurface might be obtained as
follows: let $\hv_m(\can)$ be the smallest thickening scale of $\can$
such that there exists an $m_i \in \cM \subset \cT_{\hv_m}(\can)$ with
all of $\can \subset \alex^-(m_i)$.  For any causal set $C$ pick the
\ia{} $\can_C$ with the largest value of $\hv_m(\can)$. We can then
perform the homology test on thickenings of $\can_C$. This prescription
should work reasonably for spacetimes for which there exists a scale
$V \gg V_c$ for which  the intrinsic and extrinsic curvatures are
small. 

However, it is not apriori clear that the required separation of
scales is universal. For example, if $\can$ satisfies this requirement
in $(M,g)$, then it is possible that it does not in $(M',g')$, despite
$\Phi'$ being a faithful embedding.  In such a case, our method does
not allow us to infer a relationship between the homology of $M$ and
$M'$.

In summary, a strict homology avatar of the Hauptvermutung, would
require us to pick an appropriate $\can$ without any reference to the
manifold.  Further numerical work should help us understand how stable
the homology calculation is as we varying over the \ias{} in a given
causal set \cite{numerical}. A better understanding of how both
intrinsic and extrinsic spatial geometry is encoded in a causal set
should also help resolve this question \cite{us} and would, moreover,
bring us closer to an understanding of how dynamical information is
encoded in a causal set.

\noindent {\bf Acknowledgements:} We thank M.S. Narasimhan and John
Ratcliffe for crucial help with the De Rham-Weil theorem. We also
thank Graham Brightwell, Fay Dowker, Joe Henson, Pawel Pilarczyk, and
Rafael Sorkin for discussions. We would also like to to thank the
anonymous referee for her/his remarks. Figure \ref{graph.fig} was
generated using the CHOMP homology calculator package \cite{chomp}.
DR was supported in part by the Marie Curie Research and Training Network
ENRAGE (MRTN-CT-2004-005616).

\section*{Appendix: Notation}
\begin{tabular}{||l|r||} \hline  

$C$ & Causal set  \\ \hline 
$A$ & Inextendible antichain   \\ \hline 
$\cT_{\hv}(A)$ & Thickening by $\hv$ of an antichain $A$\\ \hline 
$\cM \equiv \{m_i\}$ & Set of  maximal elements in $\cT_\hv(A)$\\ \hline 
 $P_i$ & Inclusive past set of the maximal element $m_i$ in $\cT_\hv(A)$  \\
  \hline 
$\cP\equiv \{P_i\}$ & Collection of past sets $P_i$. $\cP$ covers
  $\cT_\hv(A)$  \\ \hline 
$A_i \equiv P_i \cap A$ & Shadow of the $P_i$ onto $A$ \\ \hline 
$\bA\equiv \{A_i\}$ & Collection of shadows $A_i$. $\bA$ covers $A$\\ \hline 
$\cN(\cP), \cN(\bA)$ & Nerves associated with $\cP$ and $\bA$, respectively  \\
  \hline 
$\Psi$ & Bijection from $\cN(\cP)$ to $ \cN(\bA)$  \\ \hline 
$ \alex^\pm (x) $ & Chronological  Future/Past of $x$ \\ \hline 
$\jj^\pm(x) $ & Causal  Future/Past of $x$ \\ \hline 
$\vl $ & Volume function with respect to a Cauchy hypersurface $\Sigma$ 
\\ \hline 
$\Sigma_\vl$ & A $\vl =\mathrm{const}$ hypersurface
\\ \hline
$\xi^a $ & A future directed nowhere vanishing  
time-like vector field   \\ \hline
$\xi(s)$ & An integral curve of $\xi^a$ which intersects $\Sigma$ at
  $s$ \\ \hline 
$\xi_\vl(s)= \xi(s) \cap \Sigma_\vl$ & The intersection of $\xi(s)$
with $\Sigma_\vl$ \\ \hline  
$\sigma(p) \equiv \alex^-(p) \cap \Sigma$ & Shadow of $\alex^-(p)$
onto $\Sigma$ \\ \hline 
$\sigma_s(\vl) =  \alex^-( \xi_\vl(s)) \cap \Sigma$  & Shadow from 
an event on $\Sigma_\vl$ onto  $\Sigma$ \\ \hline
$\mS$ & Collection of shadows onto $\Sigma$ \\ \hline 
$\mS_\vl = \{ \sigma_s(\vl) \} $ & 
A  locally finite (shadow) cover  of $\Sigma $ 
\\ \hline
$\alex_i\equiv \alex(\Sigma, \xi_\vl(s_i))$ & Interval between
$\Sigma$ and $\xi_\vl(s_i)$ for fixed $\vl$ \\ \hline 
$\bI_\vl\equiv \{
\alex_i \equiv \alex(\Sigma, \xi_\vl(s_i))\} $ & Collection of
intervals $\alex_i$ \\ \hline 
$\bI$ & Collection of intervals between $\Sigma$ and events with variable $\vl$ \\
\hline 
$\cN(\bI), \cN(\mS)$ & Nerves associated with $\bI$ and $\mS$  \\
  \hline 
$\tvl $ & Convexity volume of $\Sigma$ \\ \hline 
$\al_x(\vl)$ & Volume of largest interval from  $x \in
  \Sigma$ to an event in $\Sigma_\vl$ \\ \hline 
$\vl_0 $ & Smallest  $\vl$ for which $\al_x(\vl) \gg V_c$
for all $x \in \Sigma$ \\ \hline 
$\pP_0,  \pP_1, \pP_2$ & Probabilities defined in Eqns
(\ref{pzero}), (\ref{pone}), (\ref{ptwo}) \\ \hline 
$\nN $ & Compactness Scale \\ \hline 
$\alex_{i_1\ldots i_k} \equiv \alex_{i_1} \cap \ldots \cap
\alex_{i_k}$ & Shorthand for intersection of $k$ intervals in $\bI$
\\ \hline
$P_{i_1\ldots i_k} \equiv P_{i_1} \cap \ldots \cap P_{i_k}$ & 
Shorthand for intersection of $k$ past sets in $\cP$
\\
\hline 
$(m_{s_0}, \ldots, m_{s_q} )$ & A $q$-simplex in $\cN(\bI)$, $m_{s_0}
\in \cM$ \\ \hline   
$(s_0, \ldots,  s_q) \equiv (m_{s_0}, \ldots, m_{s_q} )$ & Shorthand
for a $q$-simplex in $\cN(\cI)$ \\
\hline 
$ [s_0, \ldots, s_q] \equiv \alex_{s_0\ldots s_q}$ & The corresponding
set intersection of $q$ intervals in $\cI$\\ \hline

\end{tabular}

\end{document}